\documentclass[12pt,a4paper]{article}

\usepackage[utf8]{inputenc}
\usepackage[T1]{fontenc}
\usepackage[english]{babel}
\usepackage{mathptmx}
\usepackage[scaled=0.92]{helvet}
\usepackage{courier}
\usepackage{amssymb,amsmath,amsthm,mathtools}
\usepackage{float}
\usepackage{enumitem}
\usepackage{microtype}
\usepackage{rotating}
\usepackage{multirow}
\usepackage{hyperref}

\newtheorem{thm}{Theorem}[section]
\newtheorem{prop}[thm]{Proposition}
\newtheorem{lemma}[thm]{Lemma}

\newtheorem{@definition}[thm]{Definition}
\newenvironment{defn}{\begin{@definition}\rm}{\end{@definition}}
\newtheorem{@notation}[thm]{Notation}

\newtheorem{@example}[thm]{Example}

\newtheorem{@assumption}[thm]{Assumption}
\newenvironment{assumption}{\begin{@assumption}\rm}{\end{@assumption}}
\newtheorem{@remark}[thm]{Remark}
\newenvironment{remark}{\begin{@remark}\rm}{\end{@remark}}
\newenvironment{prf}[1][Proof]{\begin{proof}[\textsc{#1}]}{\end{proof}}

\newcommand{\rr}{\mathbb R}
\newcommand{\nn}{\mathbb N}

\newcommand{\cc}{\mathbb C}
\newcommand{\re}[1]{\mathrm{Re}\left(#1\right)}
\newcommand{\im}[1]{\mathrm{Im}\left(#1\right)}
\newcommand{\abs}[1]{\left|#1\right|}
\newcommand{\ev}[1]{E\left(#1\right)}
\newcommand{\var}[1]{\mathrm{Var}\left(#1\right)}
\newcommand{\exkurt}[1]{\mathrm{ExKurt}\left(#1\right)}
\renewcommand{\skew}[1]{\mathrm{Skew}\left(#1\right)}

\newcommand{\dy}{\,\mathrm{d}y}
\newcommand{\dz}{\,\mathrm{d}z}
\newcommand{\dt}{\,\mathrm{d}t}
\newcommand{\ds}{\,\mathrm{d}s}

\newcommand{\F}{\mathcal{F}}

\newcommand{\rir}{{}}

\newcommand{\appr}[2][{}]{\mathfrak{A}_{#1}\left(#2\right)}

\begin{document}
\title{Hedging in Lévy models and\\ the time step equivalent of jumps}
\author{
Ale\v{s} \v{C}ern\'y\footnote{
Cass Business School,
City University London,
106 Bunhill Row,
London EC1Y 8TZ, United Kingdom,
(e-mail: Ales.Cerny.1@city.ac.uk).
}
\quad Stephan Denkl\footnote{
Mathematisches Seminar,
Christian-Albrechts-Universit\"at zu Kiel,
Westring 383,
24098 Kiel, Germany,
(e-mail: denkl@math.uni-kiel.de).}
\quad Jan Kallsen\footnote{
Mathematisches Seminar,
Christian-Albrechts-Universit\"at zu Kiel,
Westring 383,
24098 Kiel, Germany,
(e-mail: kallsen@math.uni-kiel.de).}
}
\date{}
\maketitle

\begin{abstract}
We consider option hedging in a model where the underlying follows an exponential Lévy process. We derive approximations to the variance-optimal and to some suboptimal strategies as well as to their mean squared hedging errors. The results are obtained by considering the L\'evy model as a perturbation of the Black-Scholes model. The approximations depend on the first four moments of logarithmic stock returns in the Lévy model and option price sensitivities (greeks) in the limiting Black-Scholes model. We illustrate numerically that our formulas work well for a variety of Lévy models suggested in the literature.
From a theoretical point of view, it turns out that jumps have a similar effect on hedging errors as discrete-time hedging in the Black-Scholes model.\\


\end{abstract}


\section{Introduction}
A basic problem in Mathematical Finance is how the issuer of an option can hedge the resulting exposure by trading in the underlying. In complete markets, the risk can be offset completely by purchasing the replicating portfolio. In incomplete markets, however, additional criteria are necessary to determine reasonable hedging strategies. A popular approach studied intensively in the literature is \emph{variance-optimal hedging}. Here, the idea is to minimize the \emph{mean squared hedging error}, i.e., the second moment of the difference between the option's payoff and the terminal wealth of the hedging portfolio. Comprehensive overviews on the topic can be found in \cite{pham.2000, schweizer.2001}. For more recent publications, the reader is referred to \cite{cerny.kallsen.2007} and the references therein.

As a model for stock price changes, we consider exponential Lévy processes, which have been widely studied both in the theoretical and empirical literature, cf., e.g., \cite{eberlein.keller.1995, rydberg.1997, barndorff-nielsen.1995, madan.seneta.1990, madan.al.1998, raible.2000, carr.al.2002} and the monographs \cite{schoutens.2003, cont.tankov.2003}. In the context of variance-optimal hedging, \cite{hubalek.al.2006} and \cite{cerny.2007} compute semi-explicit representations of the optimal strategy and the corresponding hedging error by means of Fourier/Laplace transform methods. In addition, \cite{cerny.2007} calculates the error of the \emph{locally optimal hedge}. The related study \cite{denkl.al.2011} derives formulas for the mean squared hedging error of alternative \emph{suboptimal strategies} such as the Black-Scholes hedge, which is still prevalent in practice.

These results are exact and yield numerically tractable expressions in integral form. However, they are hard to interpret and do not allow to identify the key factors that contribute to the hedging error when deviating from the Black-Scholes model. In addition, they do not reveal how sensitively the hedging error and strategies depend on the choice of a particular parametric Lévy model.

In this study we therefore strive for reasonable first- resp.\ second-order approximations, which shed more light on the structure and dominating factors of hedging strategies and the corresponding hedging errors. It turns out that to first resp.\ second order, the Lévy process enters the solution only through its first four moments. Moreover, both strategies and hedging errors involve Black-Scholes sensitivities of the option. Depending on the payoff, the approximations are either in closed form or easy to implement numerically. In particular they bypass the need to fit return data to a specific parametric Lévy model. A numerical study shown in Section~\ref{sec:NumIll} indicates that our formulas work well for a variety of Lévy models suggested in the empirical literature. From a theoretical point of view, the approximations highlight that jumps have a similar effect on hedging errors as discrete-time hedging in a Black-Scholes environment, cf.\ Remark \ref{rem:timestep}.

In order to derive these approximations, we interpret the Lévy model at hand as a perturbed Black-Scholes model, and we compute second-order corrections that account for the perturbation. Perturbation approaches in Mathematical Finance have been considered in different contexts:

\emph{No arbitrage option pricing.} The literature on approximate option pricing is quite vast. E.g., \cite{widdicks.al.2005} expand prices in the Black-Scholes model with respect to volatility. \cite{fouque.al.2000, fouque.al.2003, alos.2006, fukasawa.2011a, fukasawa.2011b, khasminskii.yin.2005} consider expansions of option prices when the rate of mean reversion in a bivariate stochastic volatility diffusion model is fast, \cite{fukasawa.2011a, alos.2006, alos.2012} derive an expansion with respect to volatility of volatility, and \cite{antonelli.scarlatti.2009} provide a power series expansion of the price with respect to correlation. \cite{hagan.woodward.1999, benhamou.al.2009, benhamou.al.2010a, pagliarini.al.2011} consider local volatility models and derive approximate pricing formulas essentially by a Taylor expansion of the local volatility function.

\emph{Portfolio optimization and utility indifference pricing and hedging.} When considering optimal portfolio choice and consumption under transaction costs, the solution cannot be obtained explicitly even in simple models. It is typically stated in terms of quasi-variational inequalities resp.\ free boundary value problems \cite{eastham.hastings.1988, davis.norman.1990, shreve.soner.1994, morton.pliska.1995}. To shed more light on the structure of the problem, expansions with respect to the size of transactions costs were considered e.g.\ by \cite{korn.1998, janecek.shreve.2004, atkinson.wilmott.1995}. Since utility indifference option prices and hedges in the sense of \cite{hodges.neuberger.1989} are typically hard to obtain even for simple models and utility functions, first-order approximations with respect to the number of sold claims were derived as a way out \cite{mania.schweizer.2005, kramkov.sirbu.2006, kramkov.sirbu.2007}. \cite{whalley.wilmott.1997, barles.soner.1998} are early studies of 
expansions of utility indifference prices and hedges with respect to small proportional transactions costs.

\emph{Hedging errors.} Here, the literature seems to be limited to the effect of discrete-time hedging. An early contribution is \cite{toft.1996}, which studies the mean squared hedging error of the discretely implemented delta strategy in the Black-Scholes model, deriving a first-order approximation to the error with respect to the hedging interval. \cite{zhang.1999} generalizes this result to Markovian diffusion models, for which \cite{bertsimas.al.2000} consider also convergence in law of the renormalized hedging error as random variable. Extensions to irregular payoffs and more general diffusion models are to be found \cite{hayashi.mykland.2005, gobet.temam.2001, temam.2003}. \cite{geiss.2002, geiss.geiss.2006} study how the rate of convergence of the squared discretization error can be improved by using non-equidistant hedging intervals. For underlying models with jumps, \cite{tankov.voltchkova.2009} examine the rate of convergence of the discretization error of general strategies in Lévy-It\=o models. 
\cite{broden.tankov.2011} study the expected squared discretization error of variance-optimal and delta hedges in exponential Lévy models.

Our setup differs from most of the perturbation approaches above in the sense that we do not perturb a natural parameter as e.g.\ time step, number of claims, transaction costs, correlation, volatility of volatility, etc. In our situation it may not be immediately obvious in what sense an arbitrary Lévy process is to be interpreted as a one-parametric perturbation of Brownian motion. The key idea is to link the Lévy process and the Brownian motion by an appropriately chosen curve in the set of processes, parametrized by an additional artificial parameter. The only related approach in the literature that we are aware of is taken in the series of papers \cite{benhamou.al.2009, benhamou.al.2010a, benhamou.al.2010b}.

This article is organized as follows. In Section~\ref{sec:Setup} we introduce our mathematical setup. In particular, we specify the relevant hedging strategies, and we present the perturbation approach that leads to our approximate formulas. These are stated and discussed in Section~\ref{sec:Results}. For the sake of a clear exposition, all proofs are deferred to Section~\ref{sec:Proofs}. Subsequently, we illustrate our results for various parametric Lévy models. Section~\ref{sec:Conclusion} concludes.

\section{Mathematical setup} \label{sec:Setup}

For background and terminology on Lévy processes, we refer to \cite{sato.1999}. By $I$ we denote the identity process, i.e., $I_t=t$ for $t\in\rr_+$.

\subsection{Market model} \label{sec:MarketModel}
We consider a market consisting of two traded assets, a bond and a non-dividend paying stock. The price process $B$ of the bond is given by
	\begin{equation*}  B_t = e^{rt}, \quad t\in\rr_+, \end{equation*}
for a deterministic interest rate $r\geq 0$. In what follows, we will always work with discounted quantities, using $B$ as numéraire. The discounted price process $S$ of the stock is given by
	\begin{equation} \label{eq:stockPrice} S_t = S_0 e^{X_t}, \quad t\in\rr_+, \end{equation}
for a deterministic initial stock price $S_0>0$ and a real-valued Lévy process $X$ with $X_0=0$, defined on the filtered probability space $(\Omega,\F,(\F_t)_{t\in\rr_+},P)$. The filtration is assumed to be generated by $X$.

In order to carry out our analysis, we impose the following assumption on the driving Lévy process~$X$.
\begin{assumption}\label{ass:moments} We assume that
\begin{enumerate}
	\item \label{eq:expMomX} $\ev{e^{2X_1}} < \infty$,
	\item \label{eq:MomX}$\ev{\abs{X_1}^n} < \infty$ for $n\in\{1,\ldots,5\}$, and
	\item \label{eq:XnotDet} $\var{X_1} > 0$.
\end{enumerate} \end{assumption}

Given that we study second moments of hedging errors (cf.\ Section~\ref{sec:StratErr} below) and their approximation in terms of moments of $X$, Requirements~\ref{eq:expMomX} and~\ref{eq:MomX} are indispensable. The third assumption excludes the degenerate case that $S$ is deterministic.

\subsection{Option payoff function}
For the rest of the paper, we consider a fixed European contingent claim with discounted payoff $f(S_T)$ with maturity $T>0$ and (discounted) payoff function $f:\rr_+\rightarrow\rr$, which shall satisfy the following
\begin{assumption} \label{ass:LaplaceTransform} We assume that the payoff function $f:\rr_+\rightarrow\rr$ of the contingent claim under consideration is in $C^\infty(\rr_+,\rr)$ and that there exists $R\in\rr\backslash \{0\}$ with $2R\in \mathrm{int}\,D$ such that all derivatives of the mapping $x\mapsto f(e^x)e^{-Rx}$ are integrable, where
	\begin{equation*} D\coloneqq\left\{y\in\cc : \ev{\exp(\re{y}X_1}) < \infty \right\}. \end{equation*}
\end{assumption}

Depending on the Lévy process~$X$, less regularity of $f$ is needed for the proofs to work. However, for ease and clarity of exposition, we do not consider the most general statements here.

\subsection{Hedges and hedging errors} \label{sec:StratErr}
To reduce the risk arising from selling the option with payoff $f(S_T)$, we assume that the seller trades dynamically in the stock using a self-financing strategy.

\begin{defn} \label{defn:hedge}
A pair $(c,\vartheta)$ with $c\in\rr$ and a predictable $S$-integrable process $\vartheta$ is called \emph{hedge}. We refer to $c$ as the \emph{initial capital} and to $\vartheta$ as the \emph{trading strategy} of the hedge.
\end{defn}

The discounted wealth process of a hedge $(c,\vartheta)$ is $\left(c+\int_0^t \vartheta_s\, \mathrm{d}S_s\right)_{t\in[0,T]}$. We measure the performance of a hedge by its \emph{mean squared hedging error.}
\begin{defn}
The \emph{mean squared hedging error} of a hedge $(c,\vartheta)$ relative to price process $S$ is defined by
	\begin{equation*} \epsilon^2(c,\vartheta,S) \coloneqq \ev{\left(f(S_T)-c- \int_0^T\vartheta_t \, \mathrm{d}S_t\right)^2}. \end{equation*}
\end{defn}

\subsubsection{Variance-optimal hedge} \label{sec:voHedge}

In incomplete market models such as the one in Section~\ref{sec:MarketModel}, there is in general no perfect hedge that leads to a vanishing mean squared hedging error. In this situation, it is natural to look for the hedge $(v,\varphi)\in\rr\times \Theta$ with minimal mean squared hedging error, where $\Theta$ is an appropriate set of \emph{admissible trading strategies}. This approach is called \emph{variance-optimal hedging} and was studied intensely in the literature, cf., e.g., \cite{pham.2000, schweizer.2001, cerny.kallsen.2007} and the references therein. In the present context of exponential Lévy models, the set of admissible strategies is given by
	\begin{equation*} \Theta = \left\{ \vartheta \text{ predictable process} : \ev{\int_0^T \vartheta_t^2 S_{t-}^2 \mathrm{d}t} < \infty \right\}, \end{equation*}
\cite{schweizer.1994,hubalek.al.2006}.  The variance-optimal hedge $(v,\varphi)$ is determined in \cite{hubalek.al.2006}. The variance-optimal trading strategy satisfies the feedback equation
	\begin{equation} \label{eq:StrucOptStratega} \varphi_t = \xi(t,S_{t-}) + \frac{\Lambda}{S_{t-}} \left( H(t,S_{t-}) - v - \int_0^{t-} \varphi_s \mathrm{d}S_s \right), \quad t\in[0,T], \end{equation}
with deterministic functions $\xi,H:[0,T]\times \rr_+ \rightarrow\rr$ and a constant $\Lambda>0$, which are to be found in Theorem~\ref{thm:errVOexact} below. The variance-optimal initial capital $v$ is given by $v=H(0,S_0)$. Function $H$ is sometimes referred to as \emph{mean value function}. By~\eqref{eq:StrucOptStratega}, we can express the value of the variance-optimal trading strategy at time $t\in[0,T]$ as a function of the state variables $t,S_{t-},\int_0^{t-} \varphi_s \mathrm{d}S_s$, i.e.,
	\begin{equation*} \varphi_t = \varphi\left(t,S_{t-},\int_0^{t-} \varphi_s\, \mathrm{d}S_s\right), \quad t\in[0,T], \end{equation*}
where, by slight abuse of notation, the letter $\varphi$ is used to denote also the function defined as
	\begin{equation*} \varphi(t,s,g) \coloneqq \xi(t,s) + \frac{\Lambda}{s} \left( H(t,s) - v- g \right), \quad t\in[0,T], \, s\in\rr_+, \, g\in\rr. \end{equation*}
The third state variable $\int_0^{t-} \varphi_s\, \mathrm{d}S_s$ represents the past financial gains of the investor from strategy $\varphi$. For fixed $t\in[0,T]$, $s\in\rr_+$ and $g\in\rr$, we refer to $\varphi(t,s,g)$ as the \emph{variance-optimal hedge ratio in $(t,s,g)$}.

\subsubsection{Pure hedge} \label{sec:pureHedge}

For reasonable model parameters, $\Lambda$ is small and hence the contribution of the feedback term is typically modest, and it vanishes completely if $S$ is a martingale. Therefore, it makes sense to consider also the simpler \emph{pure hedge} $(v,\xi)$ defined as
	\begin{equation*}\xi_t = \xi(t,S_{t-}), \quad t\in[0,T], \end{equation*}
involving the variance-optimal initial capital $v$ and the function $\xi$ from~\eqref{eq:StrucOptStratega}. For fixed $t\in[0,T]$ and $s\in\rr_+$, we call $\xi(t,s)$ the \emph{pure hedge  ratio in $(t,s)$}. In the present Lévy setup, the trading strategy $\xi$ coincides with the so-called \emph{locally risk-minimizing hedge} in the sense of \cite{schweizer.1991}.

\subsubsection{Black-Scholes hedge} \label{sec:BShedge}

Due to its relevance in practice, we also consider the Black-Scholes hedge applied to the Lévy model of Section~\ref{sec:MarketModel}. To this end, consider a standard Brownian motion $\overline{W}$ on a filtered probability space $(\overline{\Omega}, \overline{\F}, (\overline{\F}_t)_{t\in\rr_+}, \overline{P})$, where the filtration shall be generated by $\overline{W}$. Furthermore, consider the discounted stock price process $\overline{S}$ given by
	\begin{equation} \label{eq:stockPriceBS} \overline{S}_t = S_0 e^{\mu t + \sigma \overline{W}_t}, \quad t\in\rr_+, \end{equation}
whose parameters
	\begin{equation} \label{eq:defMuSigma} \mu \coloneqq \ev{\log{\frac{S_1}{S_0}}} \quad \text{and} \quad \sigma \coloneqq \sqrt{\var{\log{\frac{S_1}{S_0}}}} \end{equation}
are chosen such that the first two moments of $\overline{X}_t \coloneqq \log\left(\frac{\overline{S}_t}{\overline{S}_0}\right)$ coincide with those of the logarithmic return process~$X$ from~\eqref{eq:stockPrice}. At time $t\in[0,T]$, the unique arbitrage-free discounted price of the contingent claim with maturity $T$ and discounted payoff $f(\overline{S}_T)$ in this model is given by $C(t,\overline{S}_t)$, where the function $C:[0,T]\times\rr_+\rightarrow\rr_+$ is given by
	\begin{equation} \label{eq:C} C(t,s) = E_{\overline{Q}}\left( \left. f(\overline{S}_T) \right| \overline{S}_t = s \right), \quad t\in[0,T], \, s\in\rr_+. \end{equation}
Here, $\overline{Q}$ denotes the unique probability measure $\overline{Q}\sim \overline{P}$ such that $\overline{S}$ is a $\overline{Q}$-martingale. Moreover, the function $C$ is continuously differentiable with respect to the second variable $s$ and
	\begin{equation*} C(0,S_0) + \int_0^T \frac{\partial}{\partial s} C(u,\overline{S}_u) \mathrm{d}\overline{S}_u = f(\overline{S}_T). \end{equation*}
Hence, $( C(0,S_0), \frac{\partial}{\partial s} C(I, \overline{S}))$ is a perfect hedge for $f(\overline{S}_T)$ in the Black-Scholes model with discounted underlying process $\overline{S}$. In the context of our Lévy model of Section~\ref{sec:MarketModel}, we use the initial capital $C(0,S_0)$ and the function $\frac{\partial}{\partial s} C$ to define a hedge $(c,\psi)$, which is given by
\begin{align}
	c &= C(0,S_0), \nonumber \\
	\psi_t &= \psi(t,S_{t-}) \label{eq:defPsi}
\end{align}
with $\psi(t,s)\coloneqq \frac{\partial}{\partial s} C(t,s).$ This hedge could e.g.\ be used by an investor who wrongly believes to trade in a Black-Scholes environment~\eqref{eq:stockPriceBS}. We refer to it as the \emph{Black-Scholes hedge applied to $S$} and to $\psi(t,s)$ as the \emph{Black-Scholes hedge ratio $(t,s)\in[0,T]\times\rr_+$}. The numerical illustration of \cite{denkl.al.2011} indicates that $(c,\psi)$ is a reasonable proxy to the variance-optimal hedge for $f\left(S_T\right)$ and exponential Lévy process $S$.

\begin{remark} \label{rem:CoinBSPVO}
If the Lévy process under consideration is Brownian motion with drift, then variance-optimal, pure and Black-Scholes hedge coincide, i.e.,
	\begin{equation*} (v,\varphi) = (v,\xi) = (c,\psi). \end{equation*}
Moreover, the mean squared hedging error of all three hedges vanishes. Finally, the mean value function coincides with the Black-Scholes pricing function in this case, i.e., $H(t,s)=C(t,s)$.
\end{remark}

\subsection{Lévy model as perturbed Black-Scholes model} \label{sec:LevyPerturbed}

\subsubsection{Outline of the approach} \label{sec:OutlineApproach}

Our goal is to derive simple and explicit approximate formulas for the initial capital, the hedge ratio and the  mean squared hedging error of the hedges from Section~\ref{sec:StratErr}. To this end, we interpret the stock price model $S$ from Section~\ref{sec:MarketModel} as a perturbed Black-Scholes model, and we compute second-order corrections that account for the perturbation. It is quite common in the Mathematical Finance literature to consider complex situations as perturbations of a simple Black-Scholes environment. Typically, the deviation from Black-Scholes is quantified in terms of a natural and usually ``small'' parameter $\lambda\in\rr$. Let us mention only three examples:
\begin{enumerate} 
	\item Option pricing and hedging in the Black-Scholes model with proportional transaction costs of size $\lambda>0$ (cf.\ \cite{whalley.wilmott.1997, barles.soner.1998}),
	\item Hedging in the Black-Scholes model at discrete points in time with distance $\lambda>0$ (cf.\ \cite{toft.1996, zhang.1999, bertsimas.al.2000, gobet.temam.2001}),
	\item Option pricing in stochastic volatility diffusion models, where the volatility has mean reversion speed $1/\lambda>0$ (cf.\ \cite{fouque.al.2000, fouque.al.2003, fukasawa.2011a, fukasawa.2011b}).
\end{enumerate}
Suppose that we are interested in a certain quantity $q(\lambda)$ of the perturbed Black-Scholes model (e.g., the indifference price and hedge under transaction costs of size $\lambda>0$, the mean squared hedging error of discrete delta hedging at time steps $\lambda>0$, or the option price under stochastic volatility with mean reversion speed $1/\lambda$ > 0). If $\lambda$ is sufficiently small, we will typically expect the first- or at least the second-order approximation
	\begin{equation} \label{eq:gLambda1} q(\lambda) \approx q(0) + q'(0)\lambda \end{equation}
resp.
	\begin{equation} \label{eq:gLambda2} q(\lambda) \approx q(0) + q'(0)\lambda + \frac{1}{2}q''(0)\lambda^2 \end{equation}
to provide a reasonable approximation. Here, $q(0)$ is the respective quantity in the Black-Scholes model itself, which is typically known explicitly. Under sufficient regularity, the approximations to $q(\lambda)$ are good if $\lambda$ is small.

In our setup, however, there is no natural small parameter that captures the deviation of the stock price process $S$ from geometric Brownian motion, and the approach \eqref{eq:gLambda1} resp.\ \eqref{eq:gLambda2} does not seem to make sense. As a way out, we introduce an artificial parameter $\lambda\in[0,1]$, where
\begin{itemize}
	\item $\lambda=1$ corresponds to the original stock price model \eqref{eq:stockPrice} of interest,
	\item $\lambda=0$ corresponds to a Black-Scholes model whose first two moments match those of the driving Lévy process in the model of interest, (\ref{eq:stockPrice}, \ref{eq:stockPriceBS}),
	\item $\lambda\in(0,1)$ corresponds to an interpolation between the two cases above, which will be specified below in Section~\ref{sec:Curve}.
\end{itemize}
Put differently, we connect the Lévy model of interest with the Black-Scholes setup via a curve in the space of Lévy processes, parametrized by $\lambda\in[0,1]$. Let now $q(\lambda)$ denote the quantity of interest in the model corresponding to parameter value $\lambda\in[0,1]$, e.g., the hedging error of the variance-optimal hedge for the option with payoff function $f$. We then suggest to use~\eqref{eq:gLambda2} as an approximation for our Lévy model of interest, i.e.\ for $\lambda=1$, which means
	\begin{equation} \label{eq:g1} q(1) \approx q(0) + q'(0) + \frac{1}{2}q''(0). \end{equation}
Both the specific form and the quality of this approximation~\eqref{eq:gLambda2} depend on the choice of the curve that connects geometric Brownian motion with~$S$ from~\eqref{eq:stockPrice}. In order to provide reasonable results, this curve should satisfy two properties.
\begin{enumerate}
	\item For the quantity $q(\lambda)$ of interest (e.g., the hedging error of the variance-optimal hedge), the derivatives $q'(0)$, $q''(0)$ need to exist and should be computable as explicitly as possible. (If the first-order approximation~\eqref{eq:gLambda1} is used, this should hold at least for $q'(0)$.)

	\item On the interval $[0,1]$, function $\lambda\mapsto q(\lambda)$ should be quadratic (resp.\ linear if \eqref{eq:gLambda1} is used) in good approximation for practically relevant cases.
\end{enumerate}

Since explicit and reasonably tight error bounds are typically hard to come by, we will assess the precision of our approximation by a collection of numerical comparisons.

\subsubsection{Curve to Brownian motion} \label{sec:Curve}
Let us now specify the curve in the space of Lévy models which connects geometric Brownian motion with the stock price process \eqref{eq:stockPrice} under consideration. We define processes $X^\lambda$ via
	\begin{equation} \label{eq:xLambda} X^\lambda_t \coloneqq \left(1-\frac{1}{\lambda}\right)\mu t + \lambda X_{\frac{t}{\lambda^2}} \quad \text{ for } \lambda\in(0,1] \text{ and } t\in\rr_+ \end{equation}
with $X$ from Section~\ref{sec:MarketModel} and $\mu$, $\sigma$ as in~\eqref{eq:mu_sigma}.
Observe that $X^\lambda$ is again a Lévy process for all $\lambda\in(0,1]$, which satisfies
	\begin{equation} \label{eq:mu_sigma} \ev{X^\lambda_t} = t \mu \quad \text{and} \quad \var{X^\lambda_t} = t\sigma^2 \quad \text{for all }\lambda\in(0,1]  \text{ and all }t\in\rr_+. \end{equation}
Equation~\eqref{eq:xLambda} does not make sense for $\lambda=0$, but we obtain Brownian motion in the limit:

\begin{lemma} \label{lem:X0}
	For $\lambda\rightarrow 0$, the family of Lévy processes $(X^\lambda)_{\lambda\in(0,1]}$ converges in law with respect to the Skorokhod topology (cf.\ \cite[VI.1]{js.03} for more details) to a Brownian motion with drift $\mu$ and volatility $\sigma$, i.e.,
	\begin{equation*} X^\lambda \xrightarrow{\mathcal{D}} \mu I + \sigma W \quad \text{as} \quad \lambda\rightarrow 0, \end{equation*}
	where $I$ denotes the identity process and $W$ is a standard Brownian motion.
\end{lemma}

We denote the limiting process by $X^0$, i.e.,
	\begin{equation} \label{eq:X0} X^0_t \coloneqq \mu t + \sigma W_t, \quad t\in\rr_+. \end{equation}

The family of Lévy processes $X^\lambda$, $\lambda\in[0,1]$, gives rise to a family of discounted stock price processes $S^\lambda$, $\lambda\in[0,1]$, namely
	\begin{equation*} S^\lambda_t \coloneqq S_0 e^{X^\lambda_t} \quad \text{for } \lambda\in[0,1] \text{ and } t\in\rr_+, \end{equation*}
where $S_0>0$ denotes the initial stock price in \eqref{eq:stockPrice}. Note that the process $S^0$ coincides in law with the Black-Scholes stock price $\overline{S}$ introduced in Section \ref{sec:BShedge}.

\subsection{Quantities to approximate} \label{sec:ListQuantities}

Our goal is to provide approximations to
\begin{enumerate}
	\item the mean value function $H(t,s)$, and in particular
	\item the initial capital $v = H(0,S_0)$ of the variance-optimal hedge from Section~\ref{sec:voHedge},
	\item the pure hedge ratio $\xi(t,s)$ from Section~\ref{sec:pureHedge},
	\item the variance-optimal hedge ratio $\varphi(t,s,g)$ from Section~\ref{sec:voHedge},
	\item the mean squared hedging error $\epsilon^2(v,\xi,S)$ of the pure hedge,
	\item the mean squared hedging error $\epsilon^2(v,\varphi,S)$ of the variance-optimal hedge,
	\item the mean squared hedging error $\epsilon^2(c,\psi,S)$ of the Black-Scholes hedge from Section~\ref{sec:BShedge}.
\end{enumerate}

In order to employ the approach outlined in Section~\ref{sec:OutlineApproach}, we have to make sure that all the above quantities are well defined.

\begin{lemma} \label{lem:QuantitiesWellDefined}
The quantities listed above are well defined in the model $S^\lambda$ for any $\lambda\in[0,1]$. Put differently, Assumptions~\ref{ass:moments} and~\ref{ass:LaplaceTransform} continue to hold and the objects from Section~\ref{sec:StratErr} are well defined if we replace $S$ by $S^\lambda$.
\end{lemma}

Let us specify the notion of \emph{second-order approximation} in our context.

\begin{defn} \label{def:approximation}
Let $Q$ denote one of the quantities listed above in the Lévy model \eqref{eq:stockPrice} of interest. Moreover, let $q(\lambda)$, $\lambda\in[0,1]$, denote the corresponding quantity with respect to $S^\lambda$, and assume that $\lambda\mapsto q(\lambda)$ is twice continuously differentiable on $[0,1]$. We call
	\begin{equation*} \appr{Q} \coloneqq \appr[0]{Q} + \appr[1]{Q} + \frac{1}{2}\appr[2]{Q} \end{equation*}
with
	\begin{equation*} \appr[0]{Q} \coloneqq q(0), \quad \appr[1]{Q} \coloneqq q'(0), \quad \appr[2]{Q} \coloneqq q''(0) \end{equation*}
\emph{second-order approximation} to $Q$.
\end{defn}

\section{Approximations to hedges and hedging errors} \label{sec:Results}

In this section, we provide the approximations in the sense of Definition~\ref{def:approximation} to the quantities listed in Section~\ref{sec:ListQuantities}. They involve two main ingredients: moments of the logarithmic return process $X=\log(S)$ and option sensitivities in the limiting Black-Scholes model $S^0$.

\subsection{Components of the approximations}
\subsubsection{Moments of the Lévy process} \label{sec:MomentRatesLogReturns}
For the first four moments of the logarithmic return process $X$ in \eqref{eq:stockPrice} we obtain
\begin{align*}
	\ev{X_t} &= \mu t,  & \var{X_t} &= \sigma^2  t, \\
	\mathrm{Skew}\left(X_t\right) &= \mathrm{Skew}\left(X_1\right) \frac{1}{\sqrt{t}}, & \mathrm{ExKurt}\left(X_t\right) &= \mathrm{ExKurt}\left(X_1\right) \frac{1}{t}.
\end{align*}
Here, $\mathrm{Skew}(Y)$ and $\mathrm{ExKurt}(Y)$ denote skewness and excess kurtosis of a random variable $Y$, i.e.,
	\begin{equation*} \skew{Y} \coloneqq \frac{\ev{(Y-\ev{Y})^3}}{\sqrt{\var{Y}}^3} \quad \text{and} \quad \exkurt{Y} \coloneqq \frac{\ev{(Y-\ev{Y})^4}}{\sqrt{\var{Y}}^4} - 3 \end{equation*}
if $\ev{Y^4}<\infty$.
Due to the scaling property in time, we refer to $\mu$, $\sigma$, $\skew{X_1}$ and $\exkurt{X_1}$ as \emph{drift}, \emph{volatility}, \emph{skewness rate} and \emph{excess kurtosis rate} of the logarithmic return process $X$.

\subsubsection{Cash greeks in the Black-Scholes model} \label{sec:CashGreeks}

Applying the reasoning of Section~\ref{sec:BShedge} to the discounted stock price process $S^0$, we see that the unique arbitrage-free discounted price at time $t\in[0,T]$ of the option with discounted payoff $f(S^0_T)$ in the model $S^0$ is given by $C(t,S^0_t)$ with function $C:[0,T]\times\rr_+\rightarrow\rr$ from~\eqref{eq:C}.

\begin{lemma} \label{lem:CisCinf}
The function $C:[0,T]\times\rr_+\rightarrow\rr$, $(t,s)\mapsto C(t,s)$, from~\eqref{eq:C} is infinitely differentiable with respect to the second variable $s$.
\end{lemma}

For $n\in\nn$ the quantity $\frac{\partial^n}{\partial s^n}C(t,S^0_t)$ represents the $n$-th order sensitivity of the option price with respect to changes in the stock price at time $t\in[0,T]$. Such sensitivities are often referred to as \emph{greeks}. Here we consider so-called \emph{cash greeks}, where the sensitivity is multiplied by the corresponding power of the stock price.

\begin{defn} \label{def:cashGreek} 
For $n\in\nn$, set
	\begin{equation*} D_n(t,s) \coloneqq s^n \frac{\partial^n}{\partial s^n} C(t,s), \quad t\in[0,T], \, s\in\rr_+, \end{equation*}
with function $C:[0,T]\times\rr_+\rightarrow\rr$, $(t,s)\mapsto C(t,s)$ from~\eqref{eq:C}.
\end{defn}

\subsection{Approximations to hedging strategies}

We begin with the approximation to the variance-optimal initial capital.
\begin{thm}[Initial capital] \label{thm:ApproxVoInitialCapital}
\begin{enumerate}[leftmargin=*]
\item The second-order approximation in the sense of Definition~\ref{def:approximation} to the mean-value function appearing in~\eqref{eq:StrucOptStratega} is
	\begin{equation*} \appr{H(t,s)} = \appr[0]{H(t,s)} + \appr[1]{H(t,s)} + \frac{1}{2} \appr[2]{H(t,s)} \end{equation*}
with
\begin{align*}
	\appr[0]{H(t,s)} &= C(t,s), \\
	\appr[1]{H(t,s)} &= \skew{X_1} \sigma^3 (T-t) \sum_{k=2}^3 a_k D_k(t,s), \\
	\appr[2]{H(t,s)} &= \skew{X_1}^2 \sigma^4 (T-t)  \left( b_2 D_2(t,s) +  \sigma^2 (T-t) \sum_{k=2}^6 c_k D_k(t,s) \right) \\ 
		& \quad {}+ \exkurt{X_1} \sigma^4 (T-t) \sum_{k=2}^4 d_k D_k(t,s),
\end{align*}
where $C$ denotes the Black-Scholes pricing function from~\eqref{eq:C}, $D_k$ are cash greeks as in Definition~\ref{def:cashGreek}, and
\begin{align*}
	a_2 &= \frac{1}{2} - \frac{1}{2} m, & a_3 &= \frac{1}{6},  & b_2 &= m - \frac{1}{6}, \\
	c_2 &= \frac{1}{2} - \frac{1}{3} m + \frac{1}{2} m^ 2, & c_3 &= \frac{13}{6} - 3 m + m^2, & c_4 &= \frac{7}{4} - \frac{3}{2} m + \frac{1}{4} m^2, \\ 
	c_5 &= \frac{5}{12} - \frac{1}{6} m , & c_6 &= \frac{1}{36}, & d_2 &= \frac{7}{12} - \frac{3}{2} m, \\
	d_3 &= \frac{1}{2} - \frac{1}{3} m, & d_4 &= \frac{1}{12}, & m &= \frac{\mu+\frac{1}{2}\sigma^2}{\sigma^2}.
\end{align*}

\item The second-order approximation of the initial capital $v$ of both variance-optimal and pure hedge is given by
	\begin{equation*} \appr{v} = \appr{H(0,S_0)}. \end{equation*}

\end{enumerate}
\end{thm}

We proceed with the approximation to the pure hedge ratio.
\begin{thm}[Pure hedge] \label{thm:ApproxRatioPureHedge}
For the second-order approximation
	\begin{equation*} \appr{\xi(t,s)} = \appr[0]{\xi(t,s)} + \appr[1]{\xi(t,s)} + \frac{1}{2}\appr[2]{\xi(t,s)} \end{equation*}
in the sense of Definition~\ref{def:approximation} to the pure hedge ratio for the option with payoff $f(S_T)$, we have
\begin{align*}
	\appr[0]{\xi(t,s)} &= \psi(t,s), \\
	\appr[1]{\xi(t,s)} &= \skew{X_1} \sigma \frac{1}{s} \left( a_2 D_2(t,s) +  \sigma^2(T-t) \sum_{k=2}^4 b_k D_k(t,s) \right), \\
	\appr[2]{\xi(t,s)} &= \skew{X_1}^2 \sigma^2 \frac{1}{s} \left(  c_2 D_2(t,s) + \sigma^2 (T-t)\sum_{k=2}^5 d_k D_k(t,s) \right) \\ 
		& \quad {}+ \skew{X_1}^2 \sigma^6(T-t)^2 \frac{1}{s} \sum_{k=2}^7 e_k D_k(t,s) \\
		& \quad {}+ \exkurt{X_1} \sigma^2 \frac{1}{s} \left( \sum_{k=2}^3 f_k D_k(t,s)+  \sigma^2 (T-t) \sum_{k=2}^5 g_k D_k(t,s)  \right),
\end{align*}
where $\psi(t,s)$ denotes the Black-Scholes hedge ratio from~\eqref{eq:defPsi}, $D_k$ are the cash greeks from Definition~\ref{def:cashGreek}, and
\begin{align*}
	a_2 &= \frac{1}{2}, & b_2 &=  1-m, & b_3 &=  1 - \frac{1}{2} m, \\
	b_4 &= \frac{1}{6}, & c_2 &= -1, & d_2 &= \frac{2}{3} + m, \\
	d_3 &= \frac{17}{6} - m, & d_4 &= \frac{3}{2} - \frac{1}{2} m, & d_5 &= \frac{1}{6}, \\
	e_2 &= 1 - 2m + m^2, & e_3 &= 7 - 10m + \frac{7}{2} m^2, & e_4 &= \frac{55}{6} -9m + 2m^2, \\
	e_5 &= \frac{23}{6} - \frac{7}{3}m + \frac{1}{4} m^2, & e_6 &= \frac{7}{12} - \frac{1}{6} m, & e_7 &= \frac{1}{36}, \\
	f_2 &= \frac{3}{2}, & f_3 &= \frac{1}{3}, & g_2 &= \frac{7}{6} - 3m, \\
	g_3 &= \frac{25}{12} - \frac{5}{2}m, & g_4 &= \frac{5}{6} - \frac{1}{3} m, & g_5 &= \frac{1}{12}, \\
	m &= \frac{\mu+\frac{1}{2}\sigma^2}{\sigma^2}. 
\end{align*}
\end{thm}

To formulate the approximation to the variance-optimal hedge ratio, we need an auxiliary result on the approximation to the mean-variance ratio $\Lambda$ from Section~\ref{sec:voHedge}.
\begin{lemma}[Mean-variance ratio] \label{lem:approxTau}
The second-order approximation to the quantity $\Lambda$ from Section~\ref{sec:voHedge} is given by
	\begin{equation*} \appr{\Lambda} = \appr[0]{\Lambda} + \appr[1]{\Lambda} + \frac{1}{2} \appr[2]{\Lambda}, \end{equation*}
\end{lemma}
where
\begin{align*}
	\appr[0]{\Lambda} &= \frac{\mu+\frac{1}{2}\sigma^2}{\sigma^2}, \\
	\appr[1]{\Lambda} &= \skew{X_1} \sigma \left( \frac{1}{6} - \appr[0]{\Lambda} \right), \\
	\appr[2]{\Lambda} &= \skew{X_1}^2 \sigma^2 \left( 2\appr[0]{\Lambda} - \frac{1}{3} \right) + \frac{1}{6} \exkurt{X_1} \sigma^2 \left( \frac{1}{2} - 7 \appr[0]{\Lambda} \right).
\end{align*}

The approximation to the variance-optimal hedge ratio is a combination of the previously obtained approximations. To this end, we write
	\begin{equation*} \varphi(t,s,g) = \xi(t,s) + \chi(t,s,g) \end{equation*}
with
	\begin{equation*} \chi(t,s,g) \coloneqq \frac{\Lambda}{s} \left( H(t,s) - v- g \right), \quad t\in[0,T], \, s\in\rr_+, \, g\in\rr, \end{equation*}
where $v$ denotes the variance-optimal initial capital, cf.\ Section~\ref{sec:voHedge}.

\begin{thm}[Variance-optimal hedge] \label{thm:ApproxRatioVoHedge}
The second-order approximation to the variance-optimal hedge ratio $\varphi(t,s,g)$ is of the form
	\begin{equation*} \appr{\varphi(t,s,g)} = \appr{\xi(t,s)} + \appr{\chi(t,s,g)}, \end{equation*}
where
	\begin{equation*} \appr{\chi(t,s,g)} = \appr[0]{\chi(t,s,g)} + \appr[1]{\chi(t,s,g)} + \frac{1}{2}\appr[2]{\chi(t,s,g)} \end{equation*}
with
\begin{align*}
	\appr[0]{\chi(t,s,g)} &= \frac{\appr[0]{\Lambda}}{s} \left( C(t,s) - C(0,S_0) - g \right), \\
	\appr[1]{\chi(t,s,g)} &= \frac{\appr[1]{\Lambda}}{s} \left( C(t,s) - C(0,S_0) - g \right) + \frac{\appr[0]{\Lambda}}{s} \left( \appr[1]{H(t,s)} - \appr[1]{v} \right),\\
	\appr[2]{\chi(t,s,g)} &= \frac{\appr[2]{\Lambda}}{s} \left( C(t,s) - C(0,S_0) - g \right) + 2 \frac{\appr[1]{\Lambda}}{s} \left( \appr[1]{H(t,s)} - \appr[1]{v} \right) \\
		& \quad {}+ \frac{\appr[0]{\Lambda}}{s} \left( \appr[2]{H(t,s)} - \appr[2]{v} \right).
\end{align*}
Here, the approximations to $\xi(t,s)$, $\Lambda$ and $H(t,s)$ are to be found in Theorem~\ref{thm:ApproxRatioPureHedge}, Lemma~\ref{lem:approxTau} and Theorem~\ref{thm:ApproxVoInitialCapital}. $C(t,s)$ denotes the Black-Scholes pricing function from~\eqref{eq:C} for the option under consideration.
\end{thm}

Observe that in the above approximations the zeroth order term is always given by the respective quantity in the limiting Black-Scholes model $S^0$. By Remark~\ref{rem:CoinBSPVO}, all three hedges under consideration coincide in the Black-Scholes case. Hence, the zeroth order approximations of initial capital and hedge ratio are given by the Black-Scholes price resp.\ the Black-Scholes hedge ratio. The second-order approximations from Theorems~\ref{thm:ApproxVoInitialCapital}, \ref{thm:ApproxRatioPureHedge} and~\ref{thm:ApproxRatioVoHedge} thus provide model-robust corrections of the Black-Scholes initial capital and the Black-Scholes hedging strategy. Our numerical study in Section~\ref{sec:NumIll} (cf.\ Tables~\ref{tab:initialCapital} and~\ref{tab:initialHedge}) shows that these corrections are excellent for a wide range of market models and payoffs.

\subsection{Approximations to hedging errors}

\begin{thm}[Variance-optimal hedging error] \label{thm:errVOapprox} The second-order approximation to the mean squared hedging error of the variance-optimal hedge $(v,\varphi)$ is given by
\begin{align*}
	\appr{\epsilon^2(v,\varphi,S)} &= \frac{1}{2}\appr[2]{\epsilon^2(v,\varphi,S)} \\ 
		& = \frac{1}{4}\sigma^4 \left(\exkurt{X_1} - \skew{X_1}^2 \right) \ev{\int_0^T e^{-\frac{(\mu+\frac{1}{2}\sigma^2)^2}{\sigma^2}(T-t)} D_2(t,S^0_t)^2 \dt}
\end{align*}
with $D_2(t,s)$ as in Definition~\ref{def:cashGreek}.
\end{thm}

\begin{thm}[Hedging error of pure hedge] \label{thm:errPureapprox} 
The second-order approximation to the mean squared hedging error of the pure hedge $(v,\xi)$ is given by
\begin{align*}
	\appr{\epsilon^2(v,\xi,S)} &= \frac{1}{2}\appr[2]{\epsilon^2(v,\xi,S)} \\
		& = \frac{1}{4}\sigma^4 \left(\exkurt{X_1} - \skew{X_1}^2 \right) \ev{\int_0^T  D_2(t,S^0_t)^2 \dt}
\end{align*}
with $D_2(t,s)$ as in Definition~\ref{def:cashGreek}.
\end{thm}

\begin{remark} \label{rem:PureApprox} 
Note that the second-order approximations to both hedging errors differ only by the exponential dampening factor $\exp\left(-(\mu+\frac{1}{2}\sigma^2)^2\sigma^{-2}(T-t)\right)$, which appears due to the feedback term of the variance-optimal trading strategy. However, its influence is typically negligible because $\ev{D_2(t,S^0_t)^2}$ is small far from maturity and dominant close to maturity. If the limiting Black-Scholes stock price process $S^0$ is a martingale, we have $\mu+\frac{1}{2}\sigma^2=0$, which implies that both approximations coincide.
\end{remark}

\begin{remark}\label{rem:timestep}
\cite{bertsimas.al.2000} study mean squared hedging errors in complete diffusion models when the replicating trading strategy of a European option is implemented discretely at time points spaced by $\Delta t$. Applied to the Black-Scholes model $S^0$, their findings yield that the mean squared hedging error $\epsilon^2(c, \psi^{\Delta}, S^0)$ of the Black-Scholes hedge $(c, \psi^{\Delta})$ implemented discretely, i.e,
	\begin{equation*} \psi^{\Delta}_t  = \psi_{\lfloor \frac{t}{\Delta t} \rfloor \Delta t}, \quad t\in[0,T], \end{equation*}
is given by
	\begin{equation*} \epsilon^2(c, \psi^{\Delta}, S^0) = \frac{1}{2} \sigma^4 \Delta t \ev{\int_0^T  D_2(t,S^0_t)^2 \dt} + o(\Delta t ) \quad \text{as } \Delta t \rightarrow 0.  \end{equation*}
Comparison with Theorem~\ref{thm:errPureapprox} suggests that, to the leading order, the risk of the pure hedge applied continuously in the Lévy model $S$ coincides with the risk from discrete delta hedging in the Black-Scholes model $S^0$ with time step
	\begin{equation*} \Delta t = \frac{1}{2} \left(\exkurt{X_1} - \skew{X_1}^2\right), \end{equation*}
which might therefore be called the \emph{time step equivalent of jumps}. E.g., taking $\skew{X_1}=\frac{0.1}{\sqrt{250}}$ and $\exkurt{X_1}=\frac{10}{250}$ as in our the numerical examples in Section~\ref{sec:NumIll}, we have 
$$\frac{1}{2} \left(\exkurt{X_1} - \skew{X_1}^2\right) \approx \frac{5}{250}.$$ 
Intuitively speaking, hedging continuously in the presence of jumps of the asset price approximately amounts to the same risk as weekly rebalanced delta hedging in a Black-Scholes market.
\end{remark}

The approximation to the hedging error of the Black-Scholes hedge applied to $S$, given by the next theorem, is a bit more involved.

\begin{thm}[Hedging error of the Black-Scholes hedge] \label{thm:errBSapprox} The second-order approximation to the mean squared hedging error of the Black-Scholes hedge $(c,\psi)$ applied to $S$, as defined in Section~\ref{sec:BShedge}, is given by
\begin{align*}
	\appr{\epsilon^2(c,\psi,S)} & = \frac{1}{2}\appr[2]{\epsilon^2(c,\psi,S)} \\ 
		& = \appr{\epsilon^2(v, \xi, S)} + \frac{1}{36} \skew{X_1}^2 \sigma^6 A(0,S_0)^2\\
		& \quad {}+ \skew{X_1}^2 \ev{\int_0^T \left(\frac{1}{2}\sigma^2 D_2(t,S^0_t) + \frac{1}{6}\sigma^4 B(t,S^0_t) \right)^2 \dt},
\end{align*}
where
\begin{align}
	A(t,s) & \coloneqq \begin{cases} (T-t)\left(  D_3(t,s) + 3D_2(t,s) \right) & \mbox{ if }\mu+\frac{1}{2}\sigma^2 = 0, \\ \frac{\widetilde{A}(t,s e^{(\mu+\frac{1}{2}\sigma^2)(T-t)})-\widetilde{A}(t,s)}{\mu+\frac{1}{2}\sigma^2} & \mbox{ if } \mu+\frac{1}{2}\sigma^2 \neq 0, \end{cases}  \label{eq:BSfuncA} \\
	B(t,s) & \coloneqq \begin{cases} (T-t) \left( D_4(t,s) + 6 D_3(t,s) + 6 D_2(t,s)\right) & \mbox{ if } \mu+\frac{1}{2}\sigma^2 = 0, \\ \frac{\widetilde{B}(t,se^{(\mu+\frac{1}{2}\sigma^2)(T-t)})-\widetilde{B}(t,s)}{\mu+\frac{1}{2}\sigma^2} & \mbox{ if } \mu+\frac{1}{2}\sigma^2 \neq 0, \end{cases} \label{eq:BSfuncB}
\end{align}
for
\begin{align}
	\widetilde{A}(t,s) & \coloneqq D_2(t,s) + D_1(t,s) - D_0(t,s), \nonumber \\
	\widetilde{B}(t,s) & \coloneqq D_3(t,s) + 3D_2(t,s). \nonumber
\end{align}
Here, $D_k(t,s)$ is as in Definition~\ref{def:cashGreek}, and the approximation $\appr{\epsilon^2(v, \xi, S)}$ to the mean squared hedging error of the pure hedge is provided by Theorem~\ref{thm:errPureapprox}.
\end{thm}

\begin{remark}
\begin{enumerate}[leftmargin=*]
	\item Theorems~\ref{thm:errPureapprox}, \ref{thm:errVOapprox} and~\ref{thm:errBSapprox} show that in general
		\begin{equation*} \appr{\epsilon^2(v,\varphi,S)} \leq \appr{\epsilon^2(v,\xi,S)} \leq \appr{\epsilon^2(c,\psi,S)},  \end{equation*}
	i.e., to the leading order the error of the pure hedge is bigger than that of the variance-optimal hedge but smaller than that of the Black-Scholes hedge. However, if the law of $X_t$ is not skewed, then the leading order errors of pure and Black-Scholes hedge coincide. As our numerical study in Section~5 shows, the difference between these two approximations is even in the case of non-zero skewness typically negligible since $\skew{X_1}^2$ is comparatively small. If, in addition to vanishing skewness, the limiting discounted Black-Scholes stock price process $S^0$ is a martingale, the approximations to the hedging errors of variance-optimal, pure and Black-Scholes hedge coincide, cf.\ Remark~\ref{rem:PureApprox}.

	\item The functions $A(t,s)$ and $B(t,s)$ in \eqref{eq:BSfuncA} and \eqref{eq:BSfuncB} are pointwise continuous in $\mu+\frac{1}{2}\sigma^2$ at $\mu+\frac{1}{2}\sigma^2=0$.
\end{enumerate}
\end{remark}

\begin{remark}
Lemma~\ref{lem:intReprCashGreeks} below provides an integral representation of cash greeks in the Black-Scholes model via the Laplace transform approach. This permits efficient evaluation of the products of cash greeks in Theorems~\ref{thm:errVOapprox}--\ref{thm:errBSapprox} by numerical integration.
\end{remark}

\section{Proofs} \label{sec:Proofs}

In this section, we present the proofs of the assertions from Sections~\ref{sec:Setup} and~\ref{sec:Results}.

\subsection{Outline}
The derivations of the approximations to initial capital, hedging ratios and hedging errors all follow the same pattern: For the respective object, we dispose of a deterministic representation in terms of an integral representation of the payoff function $f$ (stated in Section~\ref{sec:prfIntTrafo}). Formally, the quantity of interest $q(\lambda)$ relative to $S^\lambda$ can be written as
	\begin{equation*} q(\lambda) = \int h(\lambda,z) \, \mathrm{d}z. \end{equation*}
To obtain the second-order approximation to $q(1)$ in the sense of Definition~\ref{def:approximation}, we will perform three steps:
\begin{enumerate}
	\item Assure that $h(\lambda,z)$ is twice partially differentiable with respect to $\lambda$ and that integration with respect to $z$ and differentiation with respect to $\lambda$ can be interchanged.
	\item Compute $h(0,z)$, $\frac{\partial}{\partial \lambda}h(0,z)$ and $\frac{\partial^2}{\partial \lambda^2}h(0,z)$.
	\item Express the integrals over the terms in Step 2 in terms of moments of $X$ and cash greeks of the limiting Black-Scholes model $S^0$.
\end{enumerate}

\subsection{Integral representation of the payoff function} \label{sec:prfIntTrafo}

\begin{lemma} \label{lem:reprF}
There exist $R\in\rr\backslash \{0\}$ with $2R\in \mathrm{int}\,D$ and a function $p:(R+i\rr)\rightarrow\cc$ such that the payoff function~$f$ admits the representation
	\begin{equation} \label{eq:reprF} f(s) = \int_{R+i\rr} s^z \,p(z) dz. \end{equation}
Moreover, $x\mapsto \abs{R+ix}^n \abs{p(R+ix)}$ is integrable for all $n\in\nn$.
\end{lemma}

\begin{prf} By Assumption~\ref{ass:LaplaceTransform}, there exists $R\in\rr\backslash \{0\}$ with $2R\in \mathrm{int}\,D$ such that all derivatives of $l(x)\coloneqq f(e^x)e^{-Rx}$ are integrable on $\rr$. Arguing repeatedly as in the proofs of \cite[Theorem~3.3.1(f,g)]{deitmar.2005} yields that $y\mapsto y^n \mathfrak{F}l(y)$ is integrable for all $n\in\nn$, where
	\begin{equation*}\mathfrak{F}h(y) \coloneqq \int_{-\infty}^{\infty} h(x) e^{-ixy}\mathrm{d}x\end{equation*}
denotes the Fourier transform of a continuous, integrable function $h:\rr\rightarrow\rr$ in $y\in\rr$. Setting $\widetilde{p}(R+iy) \coloneqq \mathfrak{F}l(y)$, we have by the Fourier inversion Theorem (cf.\ \cite[Theorem~3.4.4]{deitmar.2005}) for all $x\in\rr$
	\begin{equation*}f(e^x) = e^{Rx}l(x) = \frac{1}{2\pi }\int_{-\infty}^\infty \mathfrak{F}l(y) e^{(R+iy)x} \,\mathrm{d}y = \frac{1}{2\pi i} \int_{R+i\rr} \widetilde{p}(z) e^{zx}\,\dz.\end{equation*}
Setting $p(R+iy) \coloneqq \frac{1}{2\pi i}\widetilde{p}(R+iy)$ and $s=e^x$ yields \eqref{eq:reprF}. Moreover, we have for all $n\in\nn$
	\begin{equation*}\abs{R+iy}^n \abs{p(R+iy)} \leq \frac{2^n}{2\pi} \max\{\abs{R}^n,\abs{y}^n\} \abs{\mathfrak{F}l(y)}. \end{equation*}
Hence, the assertion follows from the integrability of $y\mapsto y^n\mathfrak{F}l(y)$, which is stated above.
\end{prf}

From now on, we fix  $R$ as in Lemma~\ref{lem:reprF}.

\subsection{Integral representation of cash greeks in the Black-Scholes model} \label{sec:prfAsymptotics}

From the integral representation~\eqref{eq:reprF} of the payoff function, we readily obtain an integral representation for cash greeks in the Black-Scholes model $S^0$.
\begin{lemma} \label{lem:intReprCashGreeks} $D_n(t,s)$ in Definition~\ref{def:cashGreek} can be written as
	\begin{equation*} D_n(t,s) = \int_{R+i\rr} \left(\prod_{i=0}^{n-1}(z-i)\right) s^{z} e^{\frac{1}{2}\sigma^2z(z-1)(T-t)} \, p(z) \dz \end{equation*}
for any $n\in\nn$, $t\in[0,T)$ and $s\in\rr_+$.
\end{lemma}

\begin{prf}[Proof of Lemmas~\ref{lem:CisCinf} and~\ref{lem:intReprCashGreeks}]
By \cite[Lemma~5.1.2]{musiela.rutkowski.1997}, we have $S^0 = \exp\left(-\frac{1}{2}\sigma^2I + \sigma B\right)$ for a $Q$-standard Brownian motion $B$, where $Q$ denotes the unique equivalent martingale measure. Recall that the function $C:[0,T]\times\rr_+\rightarrow\rr_+$ from~\eqref{eq:C} satisfies
	\begin{equation*} C(t,s) = E_Q \left( \left. f(S^0_T) \right| S^0_t = s \right). \end{equation*}
Inserting for $f$ its integral representation from Lemma~\ref{lem:reprF} and applying Fubini's Theorem yields that
	\begin{equation*} C(t,s) = \int_{R+i\rr} s^z e^{\frac{1}{2}\sigma^2 z(z-1)(T-t)}\,p(z)\dz. \end{equation*}
Here, we used also the independence of the increments of $B$ and the fact that
	\begin{equation*} E_Q\left(e^{z\left(-\frac{1}{2}\sigma^2(T-t) + \sigma(B_T-B_t)\right)}\right) = e^{\frac{1}{2}\sigma^2z(z-1)(T-t)} \quad \text{for } z\in\cc. \end{equation*}
The application of Fubini's Theorem is justified by Lemma~\ref{lem:znExpDelta} below. By the same lemma, we have for arbitrary $n\in\nn$ that
	\begin{equation*}\int_{R+i\rr} \frac{\partial^n}{\partial s^n}\left(s^z e^{\frac{1}{2}\sigma^2 z(z-1)(T-t)}\right)p(z)\dz = \int_{R+i\rr} \left(\prod_{i=0}^{n-1}(z-i)\right)s^{z-n}e^{\frac{1}{2}\sigma^2 z(z-1)(T-t)}\,p(z)\dz\end{equation*}
is well defined for $s>0$ and $t\in[0,T)$. (Iterated) application of \cite[Satz~5.7]{elstrodt.2005} yields that we can interchange differentiation with respect to $s$ and integration with respect to $z$, which shows Lemma~\ref{lem:CisCinf} and Lemma~\ref{lem:intReprCashGreeks}.
\end{prf}

\subsection{Exact representations of hedges and hedging errors} \label{sec:ExactFormulas}

By making use of representation~\eqref{eq:reprF}, \cite{hubalek.al.2006} derive representations of variance-optimal and pure hedge and of the associated mean squared hedging errors. Their formulas are expressed in terms of the integral representation of the payoff function as in Lemma~\ref{lem:reprF} and the \emph{cumulant generating function} of the driving Lévy process.

\begin{defn} \label{def:cgf}
For $\lambda\in[0,1]$, the \emph{cumulant generating function} of $X^\lambda$ is the unique continuous function $\kappa^\lambda:D^\lambda\rightarrow\cc$ such that
	\begin{equation*} \ev{e^{z X^\lambda_t}} = e^{t\kappa^\lambda(z)} \end{equation*}
for all $t\in\rr_+$ and all $z\in D^\lambda \coloneqq \left\{ y\in\cc : \ev{e^{\re{y}X^\lambda_1}} < \infty \right\}$.
\end{defn}
For existence and uniqueness of the cumulant generating function cf.\ \cite[Lemma~7.6]{sato.1999}.

\begin{thm}[\cite{hubalek.al.2006}] \label{thm:errVOexact}
\begin{enumerate}[leftmargin=*] 
\item \label{it:OptHedge} Let $\lambda\in[0,1]$. For the stock price process~$S^\lambda$ and the contingent claim with payoff~$f(S^\lambda_T)$, the variance-optimal initial capital~$v^\lambda$ and the variance-optimal trading strategy~$\varphi^\lambda$ are given by
	\begin{equation} \label{eq:OptIc} v^\lambda=H^\lambda(0,S_0) \end{equation}
and
	\begin{equation*} \varphi^\lambda_t = \varphi^\lambda(t,S^\lambda_{t-},G^\lambda_{t-}), \quad t\in[0,T], \end{equation*}
for the function $\varphi^\lambda:[0,T]\times\rr_+\times\rr\rightarrow\rr$ given by
	\begin{equation} \label{eq:funcPhi} \varphi^\lambda(t,s,g) \coloneqq \xi^\lambda(t,s) + \frac{\Lambda^\lambda}{s}(H^\lambda(t,s)-v-g). \end{equation}
Here, the functions $H^\lambda, \xi^\lambda:[0,T]\times\rr_+\rightarrow\rr$, the process $G^\lambda$ and the constant $\Lambda^\lambda$ are defined by
\begin{align}	\bar{\kappa}^\lambda(y,z) &\coloneqq \kappa^\lambda(y+z)-\kappa^\lambda(y)-\kappa^\lambda(z), \nonumber \\
	 	\gamma^\lambda(z) & \coloneqq \frac{\bar{\kappa}^\lambda(z,1)}{\bar{\kappa}^\lambda(1,1)}, \nonumber \\
		\eta^\lambda(z) & \coloneqq \kappa^\lambda(z)-\kappa^\lambda(1)\gamma^\lambda(z), \label{eq:eta} \\
		\Lambda^\lambda & \coloneqq \frac{\kappa^\lambda(1)}{\bar{\kappa}^\lambda(1,1)}, \nonumber \\
		H^\lambda(t,s) & \coloneqq \int_{R+i\rr} s^z e^{\eta^\lambda(z)(T-t)} \, p(z) \dz, \label{eq:MeanValueProc} \\
		\xi^\lambda(t,s) & \coloneqq \int_{R+i\rr} s^{z-1} \gamma^\lambda(z) e^{\eta^\lambda(z)(T-t)} \, p(z) \dz, \label{eq:StrategyPure}\\
		G^\lambda_t & \coloneqq \int_0^t\varphi^\lambda_s \, \mathrm{d} S^\lambda_s.  \nonumber
\end{align}

\item \label{it:OptHedgeError} The corresponding mean squared hedging error of the variance-optimal hedge 
	\begin{equation*} \epsilon^2(v^\lambda,\varphi^\lambda,S^\lambda) = \ev{\left(f(S^\lambda_T)-v-\int_0^T\varphi^\lambda_t \,\mathrm{d} S^\lambda_t\right)^2} \end{equation*}
is given by
	\begin{equation*} \epsilon^2(v^\lambda,\varphi^\lambda,S^\lambda) = \int_{R+i\rr} \int_{R+i\rr} \int_0^T J_1^\lambda(t,y,z) \,p(y)p(z) \dt \dy\dz, \end{equation*}
where
\begin{align}	\rho^\lambda_j(y,z) & \coloneqq \eta^\lambda(y) + \eta^\lambda(z) - j \frac{\kappa^\lambda(1)^2}{\bar{\kappa}^\lambda(1,1)},  \quad j\in\{0,1\}, \label{eq:rho}\\
		\beta^\lambda(y,z) & \coloneqq \bar{\kappa}^\lambda(y,z) - \frac{\bar{\kappa}^\lambda(y,1)\bar{\kappa}^\lambda(z,1)}{\bar{\kappa}^\lambda(1,1)},  \nonumber\\
		J_j^\lambda(t,y,z) & \coloneqq S_0^{y+z} \beta^\lambda(y,z) e^{\kappa^\lambda(y+z)t + \rho_j^\lambda(y,z)(T-t)}, \quad  j \in \{0,1\}. \label{eq:Jd}
\end{align}

\item \label{it:PureHedgeError} The mean squared hedging error
	\begin{equation*} \epsilon^2(v^\lambda,\xi^\lambda,S^\lambda) = \ev{\left(f(S^\lambda_T)-v^\lambda-\int_0^T\xi^\lambda_t \, \mathrm{d} S^\lambda_t\right)^2} \end{equation*}
of the pure hedge (i.e., the hedge using the initial capital~$v^\lambda$ from~\eqref{eq:OptIc} and the trading strategy~$\xi^\lambda_t = \xi^\lambda(t,S^\lambda_{t-})$ from~\eqref{eq:StrategyPure}) is given by
	\begin{equation*} \epsilon^2(v^\lambda,\xi^\lambda,S^\lambda) = \int_{R+i\rr} \int_{R+i\rr} \int_0^T J_0^\lambda(t,y,z) \,p(y)p(z) \dt\dy\dz  \end{equation*}
with the function $J_0^\lambda$ as in~\eqref{eq:Jd}.
\end{enumerate}
\end{thm}

\begin{prf}
By construction, it is obvious that Assumptions~\ref{ass:moments} and~\ref{ass:LaplaceTransform} hold as well for $X^\lambda$. With the integral representation of $f$ by Lemma~\ref{lem:reprF} at hand we can apply \cite[Theorems~3.1 and~3.2]{hubalek.al.2006}, which shows Assertions~\ref{it:OptHedge} and~\ref{it:OptHedgeError}. Assertion \ref{it:PureHedgeError} follows along the same lines. Note that we can choose the same parameter $R$ for all $\lambda\in[0,1]$ because $D=D^1\subset D^\lambda$.
\end{prf}

\begin{remark}Theorem~\ref{thm:errVOexact} holds under the much milder assumption that $x\mapsto \abs{p(R+ix)}$ is integrable on $\rr$ (cf.~\cite[Section~2]{hubalek.al.2006}).\end{remark}

Following the approach of \cite{hubalek.al.2006}, \cite{denkl.al.2011} study the error of suboptimal strategies in exponential Lévy models. The next theorem restates their main result in a special case, which is sufficient for our purposes.

\begin{thm}[\cite{denkl.al.2011}] \label{thm:errBSexact} Let $\lambda\in[0,1]$. For the stock price process~$S^\lambda$, consider the initial capital $d^\lambda\in\rr$ and the trading strategy
	\begin{equation*} \vartheta^\lambda_t = \vartheta^\lambda(t,S^\lambda_{t-}), \quad t\in[0,T], \end{equation*}
with the function $\vartheta^\lambda:[0,T]\times\rr_+\rightarrow\rr$ given by
	\begin{equation}  \label{eq:psi} \vartheta^\lambda(t,s) \coloneqq \int_{R+i\rr} s^{z-1} z e^{\frac{1}{2}\nu^2z(z-1)(T-t)} \, p(z) \dz \end{equation}
for $\nu>0$. The resulting mean squared hedging error
	\begin{equation*}\epsilon^2(d^\lambda,\vartheta^\lambda,S^\lambda) = \ev{\left( f(S^\lambda_T) - d^\lambda - \int_0^T\vartheta^\lambda_t \, \mathrm{d} S^\lambda_t \right)^2}\end{equation*}
is given by
	\begin{equation}  \epsilon^2(d^\lambda,\vartheta^\lambda,S^\lambda) = (w^\lambda-d^\lambda)^2 + \int_{R+i\rr} \int_{R+i\rr}\int_0^T J^\lambda_2(t,y,z) \,p(y)p(z) \dt \dy \dz, \label{eq:errDeltaStr} \end{equation}
where
\begin{align}
	\alpha^\lambda(z,t) & \coloneqq \left(1-\kappa^\lambda(1)\int_t^T z e^{\kappa^\lambda(z)(s-T)} e^{\frac{1}{2}\nu^2z(z-1)(T-s)}  \, \ds \right) e^{\kappa^\lambda(z)(T-t)}, \label{eq:alpha} \\
	w^\lambda & \coloneqq \int_{R+i\rr} S_0^z \alpha^\lambda(z,0) \, p(z) \dz, \label{eq:w} \\
	h^\lambda(t,y,z) & \coloneqq \bar{\kappa}^\lambda(y,z)\alpha^\lambda(y,t)\alpha^\lambda(z,t) - \bar{\kappa}^\lambda(y,1)\alpha^\lambda(y,t) z e^{\frac{1}{2}\nu^2z(z-1)(T-t)} \nonumber \\
		& \; -\bar{\kappa}^\lambda(z,1)\alpha^\lambda(z,t) y e^{\frac{1}{2}\nu^2y(y-1)(T-t)} \nonumber \\ 
		& \; + \bar{\kappa}^\lambda(1,1) yz e^{(\frac{1}{2}\nu^2(z(z-1)+y(y-1))(T-t)}, \label{eq:h} \\
	J_2^\lambda(t,y,z) & \coloneqq S_0^{y+z} e^{\kappa^\lambda(y+z)t} h^\lambda(t,y,z). \label{eq:J2}
\end{align}	
\end{thm}
\begin{prf}
As noted above, Assumptions~\ref{ass:moments} and~\ref{ass:LaplaceTransform} hold as well for $X^\lambda$ by construction. With the integral representation of $f$ by Lemma~\ref{lem:reprF} at hand, we obtain that $\vartheta^\lambda$ is a $\Delta$-strategy in the sense of \cite[Definition~3.1]{denkl.al.2011}. The assertion then follows from Theorem~4.2 of the same paper. As noted in the proof of Theorem~\ref{thm:errVOexact}, we can choose the same parameter $R$ for all $\lambda\in[0,1]$.
\end{prf}

The above representations for the quantities of interest enable us to give the

\begin{prf}[Proof of Lemma~\ref{lem:QuantitiesWellDefined}]
By Theorem~\ref{thm:errVOexact}, the quantities 1--5 as well as 7 and 8 from Section~\ref{sec:ListQuantities} are obviously well defined in the model $S^\lambda$, $\lambda\in[0,1]$. From its definition and Lemma~\ref{lem:reprF}, we obtain that the Black-Scholes trading strategy applied to $S^\lambda$ (denoted by $\psi^\lambda$) admits the representation
	\begin{equation*} \psi^\lambda_t = \int_{R+i\rr} \left(S^\lambda_{t-}\right)^{z-1} z e^{\frac{1}{2}\sigma^2 z(z-1)(T-t)} \, p(z) \dz, \quad t\in[0,T]. \end{equation*}
The resulting mean squared hedging error is stated in Theorem~\ref{thm:errBSexact}.
\end{prf}

\subsection{Technicalities}

\begin{lemma} \label{lem:znExpDelta} For $n\in\nn$, $m\in\{0,1,2\}$ and $t\in[0,T)$, the mappings
	\begin{equation*}z \mapsto \abs{z^n e^{\frac{1}{2}\sigma^2z(z-1)(T-t)}} \quad \text{and} \quad z \mapsto \int_0^T \abs{z^m e^{\frac{1}{2}\sigma^2z(z-1)(T-s)}} \ds \end{equation*}
are bounded on $R+i\rr$. \end{lemma}

\begin{prf} Observe that 
	\begin{equation*} \re{\frac{1}{2}\sigma^2z(z-1)} = \frac{1}{2}\sigma^2 \left( R^2 - R - \im{z}^2 \right) = \frac{1}{2}\sigma^2(2R^2-R) - \frac{1}{2}\sigma^2\abs{z}^2 \end{equation*}
for $z\in R+i\rr$. The first assertion then follows from the fact that the mapping $x\mapsto x^n e^{-ax^2}$, $a>0$, is bounded on $\rr_+$ for all $n\in\nn$. The second assertion follows by simple integration. \end{prf}

Proposition~\ref{prop:kappa} below will be the building block for the forthcoming computations; it states the derivatives of $\kappa^\lambda(z)$ with respect to $\lambda$. To obtain these, we work with $\kappa^\lambda(z)$ in terms of its \emph{characteristic (or Lévy-Khintchine) triplet}, cf.\ \cite[Theorem~8.1]{sato.1999} for more details.

\begin{lemma} \label{lem:kappaDerivBounded}
Let $(b,c,K)$ denote the characteristic triplet of the Lévy process~$X$ with respect to the truncation function $x\mapsto 1_{[-1,1]}(x)$. Then the mapping 
	\begin{equation*}(\xi,z) \mapsto \int \abs{x^n e^{\xi z x}} \,K(dx)\end{equation*}
is bounded on $[0,1]\times (\{0,R,2R\}+i\rr)$ for $n\in\{2,\ldots,5\}$.
\end{lemma}

\begin{prf} Let $\xi\in[0,1]$ and $z\in\{0,R,2R\}+i\rr$. In the case $R\geq 0$ we have $\xi\re{z}\geq 0$, and hence
	\begin{equation} \label{eq:estKint} \int \abs{x^n e^{\xi z x}} \, K(dx) \leq \int\limits_{\{x < 0\}} \abs{x}^n\,K(dx) + \int\limits_{\{x\geq 0\}} x^n e^{2Rx} \, K(dx). \end{equation}
The first integral on the right-hand side is finite by Assumption~\ref{ass:moments}, \cite[Theorem~25.3]{sato.1999} and because $K$ is a L\'evy measure, which integrates $x\mapsto x^2$ in a neighborhood of $0$. To handle the second integral, choose $\epsilon>0$ such that $2R+\epsilon\in D$, which is possible because $2R\in \mathrm{int}\,D$ by Assumption~\ref{ass:LaplaceTransform}. Since the exponential function grows faster than any polynomial, there exists $A_\epsilon > 0$ such that $x^n e^{2Rx}\leq e^{(2R+\epsilon)x}$ for all $x\geq A_\epsilon$. Hence, we have
	\begin{equation*}\int\limits_{\{x \geq 0\}} x^n e^{2Rx}\,K(dx) \leq \int\limits_{\{0\leq x < A_\epsilon\vee 1\}} x^n e^{2Rx} \, K(dx) + \int\limits_{\{x \geq A_\epsilon \vee 1\}} e^{(2R+\epsilon)x} \, K(dx).\end{equation*}
The first integral on the right-hand side is finite since $K$ is a L\'evy measure, and the second one is finite by \cite[Theorem~25.3]{sato.1999} since $2R+\epsilon\in D$. Altogether, we have shown that both integrals in~\eqref{eq:estKint} are finite, which proves the assertion for $R\geq 0$. The case $R<0$ is treated along the same lines.
\end{prf}

\begin{prop} \label{prop:kappa}
For the family of cumulant generating functions $\kappa^\lambda(z)$ of $X^\lambda$, $\lambda\in[0,1]$, understood as a mapping $\kappa:[0,1]\times(\{0,R, 2R\}+i\rr)\rightarrow\cc$, we have the following: $\kappa$ is twice partially differentiable with respect to $\lambda$, and $\kappa$, $\frac{\partial}{\partial \lambda}\kappa$, $\frac{\partial^2}{\partial \lambda^2}\kappa$ are continuous. More specifically,
\begin{align*}
	\kappa^0(z) &= \mu z + \frac{1}{2}\sigma^2 z^2, \\		
	\frac{\partial}{\partial \lambda}\kappa^0(z) &= \frac{1}{6}\ev{(X_1-\mu)^3} z^3,\\
	\frac{\partial^2}{\partial \lambda^2}\kappa^0(z) &= \frac{1}{12} \left(\ev{ (X_1-\mu)^4} - 3\sigma^4\right) z^4.
\end{align*}
We have the estimates
	\begin{equation*} \abs{\frac{\partial^n}{\partial \lambda^n}\kappa^\lambda(z)} \leq c_1 (1+\abs{z}^{3+n}) \quad \text{for all } (\lambda,z)\in[0,1]\times \left( \{0,R,2R\}+i\rr \right), \; n\in\{0,1,2\}, \end{equation*}
where $c_1>0$ is some constant that does not depend on $\lambda$, $z$, $n$.
\end{prop}

\begin{prf}
Form the definition of $X^\lambda$ in \eqref{eq:xLambda}, it follows directly that its cumulant generating function $\kappa^\lambda$ is given in terms of $\kappa^1$ by
	\begin{equation*} \kappa^\lambda(z) = \left(1-\frac{1}{\lambda}\right)\mu z + \frac{1}{\lambda^2} \kappa^1(\lambda z), \quad \lambda\in[0,1], \, z\in\{0,R,2R\} + i\rr \subset D. \end{equation*}
For $X^0$ as defined in \eqref{eq:X0}, it is immediate that $\kappa^0(z) = \mu z + \frac{1}{2}\sigma^2 z^2$ for $z\in\{0,R,2R\} + i\rr$. Denote by $(b,c,K)$ the characteristic triplet of $X=X^1$ with respect to the truncation function $x\mapsto 1_{[-1,1]}(x)$. By \cite[Theorem~27.17]{sato.1999}, we have that
	\begin{equation*} \kappa^1(z) = bz + \frac{1}{2}cz^2 + \int \left( e^{zx} - 1 - zx 1_{[-1,1]}(x) \right) \, K(dx), \quad z\in\{0,R,2R\} + i\rr. \end{equation*}
Moreover,
	\begin{equation} \label{eq:ReprEv} \ev{X_1} = \mu = b + \int x 1_{[-1,1]^C}(x) \, K(dx) \end{equation}
by \cite[Example~25.12]{sato.1999}. Combining these two representations, we obtain that
	\begin{equation*} \kappa^\lambda(z) = \mu z + \frac{1}{2} c z^2 + \int \frac{1}{\lambda^2} \left( e^{\lambda z x} - 1 - \lambda z x \right) \, K(dx), \quad \lambda\in[0,1], \, z\in\{0,R,2R\} + i\rr. \end{equation*}
Making use of the Taylor expansion with integral remainder term
	\begin{equation*} e^{\lambda z x} = 1 + \lambda z x + \frac{1}{2}\left(\lambda z x\right)^2 + \frac{1}{2}\left(\lambda z x\right)^3 \int_0^1 e^{s\lambda z x} (1-s)^2 \,\ds, \end{equation*}
we deduce that
	\begin{equation} \label{eq:ReprKappaLambda} \kappa^\lambda(z) = \mu z + \frac{1}{2}z^2\left( c + \int x^2 \, K(dx) \right) + \frac{1}{2} z^3 \int \int_0^1 \lambda x^3 e^{s\lambda z x} (1-s)^2 \, \ds \, K(dx) \end{equation}
for $\lambda\in[0,1]$, $z\in\{0,R,2R\} + i\rr$. Observe that this representation holds also for $\lambda=0$ since by \cite[Example~25.12]{sato.1999}
	\begin{equation} \label{eq:ReprVar} \var{X_1} = \sigma^2 = c + \int x^2 \, K(dx). \end{equation}
The integrand in \eqref{eq:ReprKappaLambda} is obviously twice partially differentiable with respect to $\lambda$. Lemma~\ref{lem:kappaDerivBounded} and (iterated) application of \cite[Satz~5.7]{elstrodt.2005} yield that integration and differentiation can be interchanged. Straightforward calculations yield
\begin{align*}
	\frac{\partial}{\partial \lambda} \kappa^\lambda(z) &= \frac{1}{2} z^3 \int \int_0^1 x^3 e^{s\lambda z x} (1-s)^2 (1+\lambda z x s) \, \ds \, K(dx), \\
	\frac{\partial^2}{\partial \lambda^2} \kappa^\lambda(z) &= \frac{1}{2} z^4 \int \int_0^1 x^4 e^{s\lambda z x} s(1-s)^2 (2+\lambda z x s) \, \ds \, K(dx) 
\end{align*}
for $\lambda\in[0,1], z\in\{0,R,2R\} + i\rr$. The continuity of $\kappa$, $\frac{\partial}{\partial \lambda} \kappa$ and $\frac{\partial^2}{\partial \lambda^2} \kappa$ as well as their polynomial growth in $z$ follow now from the above representations by Lemma~\ref{lem:kappaDerivBounded} and dominated convergence. Evaluating $\frac{\partial}{\partial \lambda} \kappa^\lambda$ and $\frac{\partial^2}{\partial \lambda^2} \kappa^\lambda$ in $\lambda=0$ yields
	\begin{equation*} \frac{\partial}{\partial \lambda} \kappa^0(z) = \frac{1}{6}z^3 \int x^3 \, K(dx) \quad \text{and} \quad \frac{\partial^2}{\partial \lambda^2} \kappa^0(z) = \frac{1}{12} z^4 \int x^4 \, K(dx) \end{equation*}
for $z\in\{0,R,2R\} + i\rr$. \cite[Example~25.12]{sato.1999} derives \eqref{eq:ReprEv} and \eqref{eq:ReprVar} based on the relation between moments of a random variable and derivatives of its characteristic function, cf., e.g., \cite[Proposition~2.5(ix)]{sato.1999}. Applying the same reasoning to the higher moments of $X_1$ yields after straightforward calculations
	\begin{equation*} \int x^3 \, K(dx) = \ev{(X_1-\mu)^3} \quad \text{and} \quad \int x^4 \, K(dx) = \ev{(X_1-\mu)^4} - 3 \sigma^4. \end{equation*}
The existence of the moments is given by Assumption~\ref{ass:moments}, which completes the proof.
\end{prf}

The previous result allows us to give the proof of Lemma~\ref{lem:X0} on the convergence of $X^\lambda$ to Brownian motion as $\lambda\rightarrow 0$.

\begin{prf}[Proof of Lemma~\ref{lem:X0}]
Proposition~\ref{prop:kappa} yields directly that
	\begin{equation*} \lim_{\lambda\rightarrow 0} e^{\kappa^\lambda(iu)} = e^{\mu i u - \frac{1}{2}\sigma^2 u^2} \quad \text{for all } u\in\rr. \end{equation*}
By L\'evy's continuity theorem (cf., e.g., \cite[Proposition 2.5(vii)]{sato.1999}) the univariate marginals of $X^\lambda$ converge to the univariate marginals of $\mu I + \sigma B$ as $\lambda\rightarrow 0$, where $B$ denotes standard Brownian motion. By \cite[VII.3.6]{js.03}, this implies convergence of the whole process, which completes the proof.
\end{prf}

\begin{lemma} \label{lem:expReKappaReEta}
\begin{enumerate}[leftmargin=*]
	\item The mapping
		\begin{equation*}(\lambda,z)\mapsto \abs{e^{\kappa^\lambda(z)}} = e^{\re{\kappa^\lambda(z)}}\end{equation*}
	is bounded on $[0,1]\times(\{R,2R\}+i\rr)$.

	\item \label{it:expReEta} The mapping
		\begin{equation*}(\lambda,z)\mapsto \abs{e^{\eta^\lambda(z)}} = e^{\re{\eta^\lambda(z)}}\end{equation*}
	is bounded on $[0,1]\times(R+i\rr)$.
\end{enumerate}
\end{lemma}

\begin{prf}
\begin{enumerate}[leftmargin=*]
	\item \label{it:expKappa} For all $\lambda\in[0,1]$ and all $z\in\{R,2R\}+i\rr$ we have 
		\begin{equation} \label{eq:expReKappa} e^{\re{\kappa^\lambda(z)}} = \abs{e^{\kappa^\lambda(z)}} = \abs{\ev{e^{zX^\lambda_1}}} \leq \ev{\abs{e^{z X^\lambda_1}}} = \ev{e^{\re{z}X^\lambda_1}} = e^{\kappa^\lambda(\re{z})} \end{equation}
	by Jensen's inequality.  By Proposition~\ref{prop:kappa}, $(\lambda,r) \mapsto \kappa^\lambda(r)$ is bounded as continuous mapping on the compact set $[0,1]\times\{R, 2R\}$. Since $\re{z}\in\{R, 2R\}$, the first assertion follows.

	\item By \cite[Lemma~3.4]{hubalek.al.2006}, we have the inequality
		\begin{equation*} \abs{\gamma^\lambda(z)}^2 = \frac{\abs{\kappa^\lambda(z+1)-\kappa^\lambda(z)-\kappa^\lambda(1)}^2}{\bar{\kappa}^\lambda(1,1)^2} \leq \bar{\kappa}^\lambda(1,1) \left( \kappa^\lambda(2\re{z}) - 2 \re{\kappa^\lambda(z)} \right) \end{equation*}
	for all $\lambda\in[0,1]$ and all $z\in R+i\rr$. Hence,
	\begin{align*}
		\abs{\kappa^\lambda(1)\gamma^\lambda(z)}^2 &\leq \kappa^\lambda(1)^2\bar{\kappa}^\lambda(1,1)\kappa^\lambda(2R) - 2\kappa^\lambda(1)^2 \bar{\kappa}^\lambda(1,1)\re{\kappa^\lambda(z)} \\
			& \leq (c^\lambda)^2 + \frac{1}{4} \re{\kappa^\lambda(z)}^2 \\
			& \leq \left( c^\lambda + \frac{1}{2} \abs{\re{\kappa^\lambda(z)}} \right)^2,
	\end{align*}
	where
		\begin{equation*} c^\lambda \coloneqq \sqrt{\abs{\kappa^\lambda(1)^2\bar{\kappa}^\lambda(1,1)\kappa^\lambda(2R)} + 4 \left(2\kappa^\lambda(1)^2\bar{\kappa}^\lambda(1,1)\right)^2}, \quad \lambda\in[0,1]. \end{equation*}
	This yields
	\begin{align*}
		\re{\eta^\lambda(z)} &= \re{\kappa^\lambda(z)} - \re{\kappa^\lambda(1)\gamma^\lambda(z)} \\
			& \leq \re{\kappa^\lambda(z)} + \abs{\kappa^\lambda(1)\gamma^\lambda(z)} \\
			& \leq \re{\kappa^\lambda(z)} + \frac{1}{2} \abs{\re{\kappa^\lambda(z)}} + c^\lambda \\
			& \leq \frac{3}{2} \abs{\kappa^\lambda(R)} + c^\lambda
	\end{align*}
	because $\re{\kappa^\lambda(z)}\leq \kappa^\lambda(R)$ by~\eqref{eq:expReKappa}. Proposition~\ref{prop:kappa} yields that $\lambda\mapsto c^\lambda$ and $\lambda\mapsto \kappa^\lambda(R)$ are bounded as continuous mappings on $[0,1]$, which completes the proof.  \qedhere
\end{enumerate}
\end{prf}

\begin{lemma} \label{lem:kappaBar11}
There exists $c_2>0$ such that $\bar{\kappa}^\lambda(1,1) > c_2$ for all $\lambda\in[0,1]$.
\end{lemma}

\begin{prf}
By Assumption~\ref{ass:moments} and \eqref{eq:mu_sigma}, $\var{X^\lambda_1} = \var{X_1} > 0$ for all $\lambda\in[0,1]$. Hence,
	\begin{equation*} \var{e^{X^\lambda_1}} = \ev{e^{2X^\lambda_1}} - \ev{e^{X^\lambda_1}}^2 = e^{\kappa^\lambda(2)} - e^{2\kappa^\lambda(1)} > 0 \quad \text{for all } \lambda\in[0,1], \end{equation*}
which implies $\kappa^\lambda(2)-2\kappa^\lambda(1) = \bar{\kappa}^\lambda(1,1) > 0$ for all $\lambda\in[0,1]$. Since $\lambda\mapsto \bar{\kappa}^\lambda(1,1)$ is continuous by Proposition~\ref{prop:kappa}, it attains its minimum on $[0,1]$, which shows the assertion.
\end{prf}

\subsection{Proofs of the main theorems}

\begin{prf}[Proof of Theorem~\ref{thm:ApproxVoInitialCapital}]
By Theorem~\ref{thm:errVOexact}, we  have $v^\lambda=H^\lambda(0,S_0)$, $\lambda\in[0,1]$, for the function
	\begin{equation*} H^\lambda(t,s) = \int_{R+i\rr} s^z e^{\eta^\lambda(z)(T-t)} \, p(z) \dz, \quad t\in[0,T], \, s\in\rr_+, \end{equation*}
from~\eqref{eq:MeanValueProc}. For the remainder of the proof, we fix $t\in[0,T]$ and $s\in\rr_+$. From~\eqref{eq:eta} and Proposition~\ref{prop:kappa} we obtain
	\begin{equation} \label{eq:eta0} \eta^0(z) = \frac{1}{2} \sigma^2 z(z-1), \quad z\in R+i\rr. \end{equation}
Lemma~\ref{lem:intReprCashGreeks} then yields $H^0(t,s) = C(t,s)$ with $C(t,s)$ from~\eqref{eq:C}. To prove the assertion, we show that $\lambda\mapsto H^\lambda(t,s)$ is twice continuously differentiable on $[0,1]$, and we will identify the derivatives in $\lambda=0$. For fixed $z\in R+i\rr$, elementary calculus and Proposition~\ref{prop:kappa} yield that $\lambda\mapsto e^{\eta^\lambda(z)(T-t)}$ is twice continuously differentiable on $[0,1]$. It follows from Lemma~\ref{lem:expReKappaReEta}\eqref{it:expReEta}, Proposition~\ref{prop:kappa}, Lemma~\ref{lem:kappaBar11} and Lemma~\ref{lem:reprF} that there exists a majorant $m:(R+i\rr)\rightarrow\rr_+$ such that
	\begin{equation*} \abs{s^z \frac{\partial^n}{\partial \lambda^n} e^{\eta^\lambda(z)(T-t)}} \leq s^R m(z) \quad \text{for all } \lambda\in[0,1], \, z\in R+i\rr, \, n\in\{1,2\}, \end{equation*}
and
	\begin{equation*} \int_{-\infty}^\infty s^R m(R+ix) \abs{p(R+ix)} \, \mathrm{d}x < \infty. \end{equation*}
By (iterated) application of \cite[Satz~5.7]{elstrodt.2005} and dominated convergence, $\lambda\mapsto H^\lambda(t,s)$ is hence twice continuously differentiable on $[0,1]$, and
	\begin{equation} \label{eq:derivH} \frac{\partial^n}{\partial \lambda^n} H^\lambda(t,s) = \int_{R+i\rr} s^z \frac{\partial^n}{\partial \lambda^n} e^{\eta^\lambda(z)(T-t)} \, p(z) \dz \quad \text{for all }\lambda\in[0,1], \, n\in\{1,2\}. \end{equation}
For shorter notation, set
	\begin{equation} \label{eq:qn} q_n(z) \coloneqq \prod_{k=0}^{n-1}(z-k), \quad z\in\cc, \, n\in\nn. \end{equation}
Using the derivatives of $\kappa^\lambda(z)$ from Proposition~\ref{prop:kappa}, we obtain after lengthy but straightforward calculations that
\begin{align*}
	\left. \frac{\partial}{\partial \lambda} e^{\eta^\lambda(z)(T-t)} \right|_{\lambda=0} &= \skew{X_1} \sigma^3 (T-t) e^{\eta^0(z)(T-t)} \sum_{k=2}^3 a_k q_k(z), \\
	\left. \frac{\partial^2}{\partial \lambda^2} e^{\eta^\lambda(z)(T-t)} \right|_{\lambda=0} &=\skew{X_1}^2 \sigma^4 (T-t) e^{\eta^0(z)(T-t)}  \left( b_2 q_2(z) +  \sigma^2 (T-t) \sum_{k=2}^6 c_k q_k(z) \right) \\ 
		& \quad {}+ \exkurt{X_1} \sigma^4 (T-t) e^{\eta^0(z)(T-t)} \sum_{k=2}^4 d_k q_k(z)
\end{align*}
with constants $a_2,a_3,\ldots,d_4$ as in Theorem~\ref{thm:ApproxVoInitialCapital}. In view of \eqref{eq:eta0} and~\eqref{eq:derivH}, the assertion follows now from the integral representation of cash greeks given in Lemma~\ref{lem:intReprCashGreeks}.
\end{prf}

\begin{prf}[Proof of Theorem~\ref{thm:ApproxRatioPureHedge}]
Fix $t\in[0,T]$ and $s\in\rr_+$. By definition in~\eqref{eq:StrategyPure}, we have
	\begin{equation*} \xi^\lambda(t,s) = \int_{R+i\rr} s^{z-1} \gamma^\lambda(z) e^{\eta^\lambda(z)(T-t)} \, p(z) \dz. \end{equation*}
From Proposition~\ref{prop:kappa}, we obtain that $\gamma^0(z)=z$, and hence $\xi^0(t,s) = \frac{1}{s} D_1(t,s) = \psi(t,s)$ by~\eqref{eq:eta0} and Lemma~\ref{lem:intReprCashGreeks}. From now on, one proceeds as in the proof of Theorem~\ref{thm:ApproxRatioPureHedge}; differentiability of the integrand and existence of the majorant follow from the same lemmas. We restrict ourselves to giving the result of the essential calculation:
\begin{align*}
	\left. \frac{\partial}{\partial \lambda} \gamma^\lambda(z)e^{\eta^\lambda(z)(T-t)}  \right|_{\lambda=0} &= \skew{X_1} \sigma e^{\eta^0(z)(T-t)}  \left( a_2 q_2(z) +  \sigma^2(T-t) \sum_{k=2}^4 b_k q_k(z) \right), \\
	\left. \frac{\partial^2}{\partial \lambda^2} \gamma^\lambda(z)e^{\eta^\lambda(z)(T-t)}  \right|_{\lambda=0} &= \skew{X_1}^2 \sigma^2 e^{\eta^0(z)(T-t)} \left(  c_2 q_2(z) + \sigma^2 (T-t)\sum_{k=2}^5 d_k q_k(z) \right) \\ 
		& \quad {}+ \skew{X_1}^2 \sigma^6(T-t)^2 e^{\eta^0(z)(T-t)} \sum_{k=2}^7 e_k q_k(z) \\
		& \quad {}+ \exkurt{X_1} \sigma^2 e^{\eta^0(z)(T-t)} \sum_{k=2}^3 f_k q_k(z) \\
		& \quad {}+ \exkurt{X_1} \sigma^4 (T-t) e^{\eta^0(z)(T-t)} \sum_{k=2}^5 g_k q_k(z)
\end{align*}
for constants $a_2,\ldots,g_5$ as in Theorem~\ref{thm:ApproxRatioPureHedge}. The assertion follows now from Lemma~\ref{lem:intReprCashGreeks}.
\end{prf}

\begin{prf}[Proof of Lemma~\ref{lem:approxTau}]
The assertion follows directly by elementary calculus, using the derivatives of $\kappa^\lambda(z)$ in $\lambda=0$ stated in Proposition~\ref{prop:kappa}.
\end{prf}

\begin{prf}[Proof of Theorem~\ref{thm:ApproxRatioVoHedge}]
For fixed $t\in[0,T]$, $s\in\rr_+$ and $g\in\rr_+$, the assertion follows directly from the approximations to $\xi(t,s)$, $\Lambda$, $H(t,s)$ and $v$ given in Theorem~\ref{thm:ApproxRatioPureHedge}, Lemma~\ref{lem:approxTau} and Theorem~\ref{thm:ApproxVoInitialCapital}.
\end{prf}

\begin{prf}[Proof of Theorems~\ref{thm:errVOapprox} and~\ref{thm:errPureapprox}]
For shorter notation, let $\epsilon_0^2(\lambda)\coloneqq \epsilon^2(v^\lambda,\xi^\lambda,S^\lambda)$ and $\epsilon_1^2(\lambda)\coloneqq \epsilon^2(v^\lambda,\varphi^\lambda,S^\lambda)$, $\lambda\in[0,1]$, be the mean squared hedging errors of pure and variance-optimal hedge in the model $S^\lambda$. In order to prove the assertion, we will show that $\lambda\mapsto \epsilon^2_j(\lambda)$, $j\in\{0,1\}$, is twice continuously differentiable on $[0,1]$, and we will identify the derivatives in $\lambda=0$. To this end, we use the deterministic representation of $\epsilon^2_j(\lambda)$ by Theorem~\ref{thm:errVOexact}. Inserting $\kappa^0$ from Proposition~\ref{prop:kappa} immediately yields that $\epsilon^2_0(0) = \epsilon^2_1(0) = 0$. For fixed $(t,y,z)\in[0,T]\times(R+i\rr)\times(R+i\rr)$, the mapping $\lambda\mapsto J^\lambda_j(t,y,z)$ with $J^\lambda$ from \eqref{eq:Jd} is twice continuously differentiable on $[0,1]$ by Proposition~\ref{prop:kappa} and elementary differential calculus. 
Moreover, by Proposition~\ref{prop:kappa}, Lemmas~\ref{lem:reprF}, \ref{lem:expReKappaReEta} and \ref{lem:kappaBar11} $\frac{\partial}{\partial \lambda} J^\lambda_j(t,y,z)$ and $\frac{\partial^2}{\partial \lambda^2} J^\lambda_j(t,y,z)$ admit a majorant $m:[0,T]\times(R+i\rr)\times(R+i\rr)\rightarrow\rr_+$, more precisely,
	\begin{equation*} \abs{\frac{\partial^n}{\partial \lambda^n} J^\lambda_j(t,y,z) } \leq m(t,y,z) \quad \text{for all } \lambda\in[0,1], \; t\in[0,T], \; y,z\in R+i\rr, \; n\in\{1,2\} \end{equation*}
such that
	\begin{equation*} \int_{-\infty}^\infty \int_{-\infty}^\infty \int_0^T m(t,R+iu,R+iv) \abs{p(R+iu)} \abs{p(R+iv)} \, \dt\, \mathrm{d}u\, \mathrm{d}v < \infty. \end{equation*}
Hence, by (iterated) application of \cite[Satz~5.7]{elstrodt.2005} and dominated convergence, $\lambda\mapsto \epsilon^2_j(\lambda)$ is twice continuously differentiable on $[0,1]$ and
	\begin{equation*} \frac{\partial^n}{\partial \lambda^n} \epsilon^2_j(\lambda) =  \int_{R+i\rr} \int_{R+i\rr}\int_0^T \frac{\partial^n}{\partial \lambda^n} J^\lambda_j(t,y,z) \,p(y)p(z) \dt \dy \dz \quad \text{for all } \lambda\in[0,1], \; n\in\{1,2\}. \end{equation*}
By lengthy but straightforward calculations, we obtain from Proposition~4.9 that we have $\frac{\partial}{\partial \lambda} \left. J^\lambda_j(t,y,z)\right|_{\lambda=0} = 0$, $j\in\{0,1\}$, and
\begin{align*} \frac{\partial^2}{\partial \lambda^2} \left. J^\lambda_j(t,y,z) \right|_{\lambda=0} &= \frac{1}{2} \sigma^4 \left(\exkurt{X_1}- \skew{X_1}^2\right) 		e^{-j\frac{\left(\mu+\frac{1}{2}\sigma^2\right)^2}{\sigma^2}(T-t)} \\ 
	& \quad {}\times S_0^{y+z}  e^{\kappa^0(y+z)t} e^{\eta^0(y)(T-t)}e^{\eta^0(z)(T-t)} y(y-1)z(z-1) .
\end{align*}
In order to interpret the integral over this derivative in the desired way, we use Fubini's Theorem (whose application is justified by Lemmas~\ref{lem:reprF} and \ref{lem:expReKappaReEta}) and Lemma~\ref{lem:intReprCashGreeks} in order to obtain
\begin{align*} 
	&\int\limits_\rir \int\limits_\rir \int_0^T S_0^{y+z} e^{\kappa^0(y+z)t} e^{(\eta^0(y)+\eta^0(z))(T-t)} y(y-1) z(z-1) \, p(y)p(z) \dt \dy\dz \\
	&= \int_0^T \int\limits_\rir \int\limits_\rir \ev{(S^0_t)^{y+z}} e^{(\eta^0(y)+\eta^0(z))(T-t)} y(y-1) z(z-1) \,p(y)p(z) \dy\dz \dt \\
	&= \ev{ \int_0^T \left( (S^0_t)^2 \int\limits_\rir (S^0_t)^{y-2} e^{\eta^0(y)(T-t)}y(y-1) \,p(y)\dy \right)^2  \dt} \\ 
	&= \ev{\int_0^T D_2(t,S^0_t)^2 \dt},
\end{align*}
which completes the proof.
\end{prf}

\begin{prf}[Proof of Theorem~\ref{thm:errBSapprox}]
By definition in Section~\ref{sec:BShedge} and by Lemma~\ref{lem:intReprCashGreeks}, the Black-Scholes hedge $(c^\lambda,\psi^\lambda)$ applied to $S^\lambda$ admits the integral representation
\begin{align*}
	c^\lambda &= \int_{R+i\rr} S_0^z e^{\frac{1}{2}\sigma^2z(z-1)T} p(z) \dz, \\
	\psi^\lambda_t &= \int_{R+i\rr} S_{t-}^{z-1} z e^{\frac{1}{2}\sigma^2z(z-1)(T-t)} p(z) \dz, \quad t\in[0, T].
\end{align*}
Hence, Theorem~\ref{thm:errBSexact} can be applied with $\nu=\sigma$ and $d^\lambda=c^\lambda$. Thus, we obtain a deterministic integral representation of the hedging error $\epsilon^2(c^\lambda,\psi^\lambda,S^\lambda)$ for $\lambda\in[0,1]$. Observe that $\epsilon^2(c^0, \psi^0, S^0)=0$. The reasoning to show the assertion is now analogous to the proof of Theorems~\ref{thm:errVOapprox} and~\ref{thm:errPureapprox}. The existence of the necessary majorants and differentiability of $\lambda\mapsto \epsilon^2(c^\lambda,\psi^\lambda,S^\lambda)$ on $[0,1]$ follow from the same lemmas. Tedious but straightforward calculations based on Proposition~\ref{prop:kappa} yield $\frac{\partial}{\partial \lambda} \left. J_2^\lambda(t,y,z) \right|_{\lambda=0} = 0$ and
\begin{align}
	\frac{\partial^2}{\partial \lambda^2} \left. J_2^\lambda(t,y,z) \right|_{\lambda=0} & = S_0^{y+z} e^{\kappa^0(y+z)t} e^{(\eta^0(y)+\eta^0(z))(T-t)}  \nonumber \\
			& \quad \times \left( \frac{1}{2} \sigma^4 \exkurt{X_1} y(y-1)z(z-1) + \frac{1}{18}\sigma^8 \skew{X_1}^2 b(y,t) b(z,t)  \right. \nonumber \\
			& \quad \quad {}+ \frac{1}{6} \sigma^6 \skew{X_1}^2 \left( y(y-1)b(z,t) + z(z-1)b(y,t) \right) \bigg) \label{eq:expJ2}
\end{align}
for $J_2^\lambda$ from \eqref{eq:J2} in the case $(d^\lambda,\vartheta^\lambda)=(c^\lambda,\psi^\lambda)$ and with
	\begin{equation*} c(z) \coloneqq \left(\mu+\frac{1}{2}\sigma^2\right)z \quad \text{and} \quad b(z,t) \coloneqq (z^4-z^2) \int_t^T e^{c(z)(s-t)}, \quad z\in R+i\rr, \; t\in[0,T]. \end{equation*}
In the case $\mu+\frac{1}{2}\sigma^2=0$, we have $b(z,t) = (T-t) (q_4(z)+6q_3(z)+6q_2(z))$ with $q_n(z)$ as defined in~\eqref{eq:qn}. The expected time integral in the assertion is obtained as in the proof of Theorems~\ref{thm:errVOapprox} and \ref{thm:errPureapprox}. In the case $\mu+\frac{1}{2}\sigma^2 \neq 0$, we have $b(z,t) = (q_3(z)+3q_2(z)) \left( \exp(c(z)(T-t))-1\right) \left( \mu+\frac{1}{2}\sigma^2 \right)^{-1}$. To see how to handle the additional term $\exp(c(z)(T-t))$, let us exemplarily consider the relevant part of the second summand in~\eqref{eq:expJ2}
	\begin{multline*} \int \int \int_0^T S_0^{y+z} e^{\kappa^0(y+z)t} e^{(\eta^0(y)+\eta^0(z))(T-t)} b(y,t)b(z,t) \, p(y)p(z)\dt\dy\dz \\
	= \ev{\int_0^T \left( \int \left(S^0_t\right)^z e^{\eta^0(z)(T-t)} \frac{e^{c(z)(T-t)}-1}{\mu+\frac{1}{2}\sigma^2} (q_3(z)+3q_2(z)) \,p(z)\dz \right)^2 \dt}, \end{multline*}
where we used that $S_0^{y+z}e^{\kappa^0(y+z)t} = \ev{(S^0_t)^{y+z}}$ and Fubini's Theorem, whose application is justified by Lemmas~\ref{lem:reprF} and \ref{lem:expReKappaReEta}. By Lemma~\ref{lem:intReprCashGreeks}, we obtain that
\begin{multline*}
	\int \left(S^0_t\right)^z e^{\eta^0(z)(T-t)} \frac{e^{c(z)(T-t)}-1}{\mu+\frac{1}{2}\sigma^2} (q_3(z)+3q_2(z))  \,p(z)\dz\\
	\shoveleft{= \int e^{\eta^0(z)(T-t)} \frac{ \left(S^0_t e^{(\mu+\frac{1}{2}\sigma^2)(T-t)}\right)^z - \left(S^0_t\right)^z }{\mu+\frac{1}{2}\sigma^2} \left( q_3(z)+3q_2(z) \right) \,p(z)\dz } \\
	= \frac{ D_3(t,S^0_t e^{(\mu+\frac{1}{2}\sigma^2)(T-t)}) + 3D_2(t,S^0_t e^{(\mu+\frac{1}{2}\sigma^2)(T-t)}) - D_3(t, S^0_t) - 3D_2(t, S^0_t)}{\mu+\frac{1}{2}\sigma^2}.
\end{multline*}
Differentiability and the interpretation of the derivatives in $\lambda=0$ of the mapping $\lambda\mapsto \left(w^\lambda-d^\lambda\right)$ are treated completely analogously to the proof of Theorem~\ref{thm:ApproxVoInitialCapital}. Summing up all calculations, we obtain
	\begin{equation*} \left. \epsilon^2(c^\lambda,\psi^\lambda,S^\lambda)\right|_{\lambda=0} = \left.\frac{\partial}{\partial\lambda}\epsilon(c^\lambda,\psi^\lambda,S^\lambda) \right|_{\lambda=0} = 0 \end{equation*}
and
\begin{align*}
	\left.\frac{\partial^2}{\partial\lambda^2}\epsilon(c^\lambda,\psi^\lambda,S^\lambda) \right|_{\lambda=0} & = \frac{1}{18} \skew{X_1}^2 \sigma^6 A(0,S_0)^2 \\
		& \quad {}+ \frac{1}{2} \exkurt{X_1} \sigma^4 \ev{\int_0^T D_2(t,S^0_t)^2 \dt} \\
		& \quad {}+ \frac{1}{18}\skew{X_1}^2 \sigma^8 \ev{\int_0^T B(t,S^0_t)^2 \dt} \\
		& \quad {}+ \frac{1}{3}\skew{X_1}^2 \sigma^6 \ev{\int_0^T B(t,S^0_t) D_2(t,S^0_t) \dt}
\end{align*}
with the mappings $A$, $B:[0,T]\times\rr_+\rightarrow\rr$ as defined in~\eqref{eq:BSfuncA}, \eqref{eq:BSfuncB}. Reordering and comparison with Theorem~\ref{thm:errPureapprox} completes the proof.
\end{prf}

\section{Numerical comparison} \label{sec:NumIll}

In this section, we examine the accuracy of the approximations from Section~\ref{sec:Results} by numerical examples. To this end, we compare exact and approximate initial capital, initial hedge ratio and root mean squared hedging error of the variance-optimal hedge. Moreover, we compare the exact and approximate root mean squared hedging error of the Black-Scholes hedge. We perform our study for European call options in three different parametric Lévy models.

\subsection{Market models} \label{sec:NumMarketModels}
As parametric market models for the discounted stock, we consider Merton's jump-diffusion (JD) model with normal jumps \cite{merton.1976}, the normal inverse Gaussian (NIG) model \cite{barndorff-nielsen.1998} and the variance gamma (VG) model \cite{madan.seneta.1990} for various parameter choices.

As initial stock price, we always set $S_0=100$. Moreover, we fix the parameters of all models such that
\begin{align*}
	\mu = \ev{\log\left(X_1\right)} &= -0.08, \\
	\sigma^2 = \var{\log\left(X_1\right)} &= 0.4^2, \\
	\mathrm{Skew}\left(\log\left(X_1\right)\right) &= \frac{0.1}{\sqrt{250}}.
\end{align*}
The excess kurtosis rate $\mathrm{ExKurt}\left(\log\left(X_1\right)\right)$ is chosen as $2/250$, $5/250$, $10/250$, respectively. All these choices are well within the range of empirically plausible values, cf., e.g., \cite[Table~4]{carr.al.2002}. Note that skewness rate and excess kurtosis rate are reported such that one directly recovers the values on a daily basis, assuming $250$ trading days per year. Moreover, our choice is such that $\mu+\frac{1}{2}\sigma^2=0$, i.e.\ the stock has the risk-free rate of return. Hence, the mean squared hedging error of variance-optimal and pure hedge coincide in this situation, cf.\ Remark~\ref{rem:PureApprox}.

NIG and VG are models with four parameters, and so the specification of the first four moments of logarithmic returns leaves no degree of freedom. The JD model, however, has five parameters. In order to eliminate the additional degree of freedom, the parameters are chosen such that the volatility arising from the jump component explains $70\%$ of the overall volatility of logarithmic returns.

In order to calculate the exact values of the quantities of interest, we use the formulas from Section~\ref{sec:ExactFormulas} and perform standard numerical quadrature.

\subsection{Option payoff function}
We consider European calls with strike $K=95$, $100$ or $105$, respectively, and maturity $T=1/12$, $1/4$ or $1/2$, measured in years. The corresponding payoff function $f(s) = (s-K)^+$ allows for an integral representation as in~\eqref{eq:reprF}, given by
	\begin{equation*} f(s) = \frac{1}{2\pi i} \int_{R-i\infty}^{R+i\infty} s^z \frac{K^{1-z}}{z(z-1)} \, dz \end{equation*}
for arbitrary $R>1$, cf.\ \cite[Lemma~4.1]{hubalek.al.2006}. Strictly speaking, the kinked payoff function of the European call does not meet the smoothness requirement of Assumption~\ref{ass:LaplaceTransform}. Nevertheless, the approximate formulas from Section~\ref{sec:Results} are well defined in this situation, as one easily shows by use of Lemma~\ref{lem:znExpDelta}. Hence, we can and will use them in our numerical comparison.

\subsection{Hedges and hedging errors}
In any of the above cases, we compute the initial capital~$v$, the initial hedge ratio $\varphi(0, S_0, v)$ and the square root $\sqrt{\epsilon^2(v,\varphi,S)}$ of the mean squared hedging error of the variance-optimal hedge. These are compared to the respective approximations from Theorems~\ref{thm:ApproxVoInitialCapital}, \ref{thm:ApproxRatioVoHedge} and~\ref{thm:errVOapprox}. Moreover, we report the corresponding Black-Scholes price $c=C(0,S_0)$ and the initial Black-Scholes hedge ratio $\psi(0,S_0)$. Finally, we compute the square root $\sqrt{\epsilon(c,\psi,S)}$ of the exact mean squared hedging error of the Black-Scholes hedge and compare it to the approximation from Theorem~\ref{thm:errBSapprox}.

\subsection{Discussion of the numerical results}
Table~\ref{tab:initialCapital} shows the exact and approximate variance-optimal initial capital as well as the Black-Scholes price for different models and payoffs. Table~\ref{tab:initialHedge} reports the exact and approximate variance-optimal hedge ratio for $t=0$ as well as the initial Black-Scholes hedge ratio. For both quantities, the exact values mostly coincide across models, and the approximation is precise up to the last digit. For high excess kurtosis and short maturity, the performance of the approximations is slightly worse, but also the improvement compared to the mere Black-Scholes value becomes more pronounced.

Table~\ref{tab:hedgingError} shows exact and approximate values for the square root of the mean squared hedging error of the variance-optimal hedge. In brackets we report the exact resp.\ approximate square root of the mean squared hedging error of the Black-Scholes hedge. We observe that the difference between the approximations to both strategies seems negligible for practical purposes. Moreover, the approximations tend to slightly overestimate the exact values. The performance becomes worse for shorter time to maturity and higher excess kurtosis. In the case of the variance-optimal hedge, e.g.\ for $K=100$ and $\mathrm{ExKurt(X_1)}=2/250$, the relative deviation of the approximate value from the average exact value over all models amounts to $6.7\%$ for $T=1/12$ and to $2.4\%$ for $T=1/2$. For $K=100$ and $\mathrm{ExKurt(X_1)}=10/250$, the relative deviation accounts for $18\%$ in the case $T=1/12$ and for $6.0\%$ in the case $T=1/2$. As already pointed out in \cite{denkl.al.2011}, we finally see from 
the respective hedging errors that the mere Black-Scholes hedge is a satisfying proxy to the variance-optimal hedge.

As mentioned above, the approximations to variance-optimal and pure hedge from Theorems~\ref{thm:errVOapprox} and~\ref{thm:errPureapprox} coincide in our study since we choose $\mu+\frac{1}{2}\sigma^2=0$. However, numerical experiments that are not shown here indicate that, for typical parameter choices, the difference between these two approximations is negligible (in the magnitude of less than $1\%$) also if $\mu+\frac{1}{2}\sigma^2\neq 0$. Hence, for practical purposes the most simple of our formulas -- the one from Theorem~\ref{thm:errPureapprox} -- should be used to approximately quantify the error of either pure, variance-optimal or Black-Scholes hedge.

\begin{sidewaystable} 
{\small
\begin{tabular}{cc|ccccc|ccccc|ccccc}
\multirow{3}{*}{$\mathrm{ExKurt(X_1)}$} & \multirow{3}{*}{$K$} &&& $T=\frac{1}{12}$ &&&&& $T=\frac{1}{4}$ &&&&& $T=\frac{1}{2}$ && \\
&&&&&&&&&&&&&&&& \\
{} & {} & JD & NIG & VG & BS & Approx & JD & NIG & VG & BS & Approx & JD & NIG & VG & BS & Approx \\
\hline
&&&&&&&&&&&&&&&&\\
{} & 95 & 7.406 & 7.406 & 7.406 & 7.424 & 7.406 & 10.511 & 10.511 & 10.511 & 10.520 & 10.511 & 13.641 & 13.641 & 13.641 & 13.644 & 13.641 \\
$\frac{2}{250}$ & 100 & 4.589 & 4.589 & 4.589 & 4.604 & 4.589 & 7.961 & 7.961 & 7.961 & 7.966 & 7.961 & 11.247 & 11.247 & 11.247 & 11.246 & 11.247 \\
{} & 105 & 2.631 & 2.631 & 2.631 & 2.634 & 2.631 & 5.907 & 5.907 & 5.907 & 5.906 & 5.907 & 9.202 & 9.202 & 9.202 & 9.197 & 9.202 \\
&&&&&&&&&&&&&&&& \\
{} & 95 & 7.385 & 7.386 & 7.385 & 7.424 & 7.384 & 10.496 & 10.496 & 10.496 & 10.520 & 10.496 & 13.631 & 13.631 & 13.631 & 13.644 & 13.631 \\
$\frac{5}{250}$ & 100 & 4.562 & 4.564 & 4.562 & 4.604 & 4.562 & 7.946 & 7.946 & 7.946 & 7.966 & 7.946 & 11.237 & 11.237 & 11.237 & 11.246 & 11.237 \\
{} & 105 & 2.614 & 2.614 & 2.614 & 2.634 & 2.613 & 5.894 & 5.895 & 5.894 & 5.906 & 5.894 & 9.194 & 9.194 & 9.194 & 9.197 & 9.194 \\
&&&&&&&&&&&&&&&& \\
{} & 95 & 7.351 & 7.355 & 7.351 & 7.424 & 7.348 & 10.472 & 10.473 & 10.472 & 10.520 & 10.472 & 13.614 & 13.615 & 13.614 & 13.644 & 13.614 \\
$\frac{10}{250}$ & 100 & 4.520 & 4.524 & 4.518 & 4.604 & 4.517 & 7.922 & 7.923 & 7.922 & 7.966 & 7.921 & 11.221 & 11.222 & 11.221 & 11.246 & 11.221 \\
{} & 105 & 2.586 & 2.588 & 2.586 & 2.634 & 2.583 & 5.874 & 5.875 & 5.874 & 5.906 & 5.874 & 9.180 & 9.180 & 9.180 & 9.197 & 9.180
 
\end{tabular}
\caption{Exact and approximate variance-optimal initial capital and Black-Scholes price for $\mu=-0.08$, $\sigma = 0.4$, $\mathrm{Skew(X_1)}=\frac{0.1}{\sqrt{250}}$ and varying excess kurtosis $\mathrm{Exkurt(X_1)}$, strike $K$ and maturity $T$}
\label{tab:initialCapital}
}
\end{sidewaystable}

\begin{sidewaystable}
{\small
\begin{tabular}{cc|ccccc|ccccc|ccccc}
\multirow{3}{*}{$\mathrm{ExKurt(X_1)}$} & \multirow{3}{*}{$K$} &&& $T=\frac{1}{12}$ &&&&& $T=\frac{1}{4}$ &&&&& $T=\frac{1}{2}$ && \\
&&&&&&&&&&&&&&&& \\
{} & {} & JD & NIG & VG & BS & Approx & JD & NIG & VG & BS & Approx & JD & NIG & VG & BS & Approx \\
\hline
&&&&&&&&&&&&&&&&\\
{} & 95 & 0.696 & 0.696 & 0.696 & 0.692 & 0.696 & 0.642 & 0.642 & 0.642 & 0.639 & 0.642 & 0.628 & 0.628 & 0.628 & 0.627 & 0.628 \\
$\frac{2}{250}$ & 100 & 0.528 & 0.528 & 0.528 & 0.523 & 0.528 & 0.543 & 0.543 & 0.543 & 0.540 & 0.543 & 0.558 & 0.558 & 0.558 & 0.556 & 0.558 \\
{} & 105 & 0.363 & 0.363 & 0.363 & 0.358 & 0.363 & 0.446 & 0.446 & 0.446 & 0.443 & 0.446 & 0.490 & 0.490 & 0.490 & 0.488 & 0.490 \\
&&&&&&&&&&&&&&&& \\
{} & 95 & 0.697 & 0.697 & 0.697 & 0.692 & 0.697 & 0.643 & 0.643 & 0.643 & 0.639 & 0.643 & 0.629 & 0.629 & 0.629 & 0.627 & 0.629 \\
$\frac{5}{250}$ & 100 & 0.530 & 0.530 & 0.530 & 0.523 & 0.530 & 0.544 & 0.544 & 0.544 & 0.540 & 0.544 & 0.559 & 0.559 & 0.559 & 0.556 & 0.559 \\
{} & 105 & 0.367 & 0.367 & 0.367 & 0.358 & 0.367 & 0.447 & 0.447 & 0.447 & 0.443 & 0.448 & 0.491 & 0.491 & 0.491 & 0.488 & 0.491 \\
&&&&&&&&&&&&&&&& \\
{} & 95 & 0.699 & 0.699 & 0.699 & 0.692 & 0.699 & 0.645 & 0.645 & 0.645 & 0.639 & 0.645 & 0.631 & 0.631 & 0.631 & 0.627 & 0.631 \\
$\frac{10}{250}$ & 100 & 0.534 & 0.534 & 0.534 & 0.523 & 0.535 & 0.546 & 0.546 & 0.546 & 0.540 & 0.547 & 0.561 & 0.561 & 0.561 & 0.556 & 0.561 \\
{} & 105 & 0.373 & 0.372 & 0.373 & 0.358 & 0.373 & 0.450 & 0.450 & 0.450 & 0.443 & 0.450 & 0.493 & 0.493 & 0.493 & 0.488 & 0.493
 
\end{tabular}
\caption{Exact and approximate initial variance-optimal hedge ratio $\varphi(0,S_0,v)$ as well as initial Black-Scholes hedge ratio $\psi(0,S_0)$ for $\mu=-0.08$, $\sigma = 0.4$, $\mathrm{Skew(X_1)}=\frac{0.1}{\sqrt{250}}$ and varying excess kurtosis $\mathrm{Exkurt(X_1)}$, strike $K$ and maturity $T$}
\label{tab:initialHedge}
}
\end{sidewaystable}

\begin{sidewaystable} 
{\small
\begin{tabular}{cc|cccc|cccc|cccc}
\multirow{3}{*}{$\mathrm{ExKurt(X_1)}$} & \multirow{3}{*}{$K$} & \multicolumn{4}{|c|}{$T=\frac{1}{12}$} & \multicolumn{4}{|c|}{$T=\frac{1}{4}$} & \multicolumn{4}{|c}{$T=\frac{1}{2}$}\\
&&&&&&&&&&&&& \\
{} & {} & JD & NIG & VG & Approx & JD & NIG & VG & Approx & JD & NIG & VG & Approx \\
\hline
&&&&&&&&&&&&&\\
\multirow{6}{*}{$\frac{2}{250}$} & \multirow{2}{*}{95} & 0.756 & 0.746 & 0.760 & 0.808 & 0.812 & 0.807 & 0.814  & 0.841 & 0.827 & 0.824 & 0.829 & 0.847 \\
{} &  & (0.764) & (0.753) & (0.768) & (0.810) & (0.817) & (0.811) & (0.819)  & (0.843) & (0.830) & (0.827) & (0.832) & (0.849) \\
 & \multirow{2}{*}{100} & 0.837 & 0.828 & 0.840 & 0.891 & 0.859 & 0.855 & 0.861 & 0.889 & 0.865 & 0.864 & 0.868 & 0.887 \\
 &  & (0.846) & (0.836) & (0.851) & (0.893) & (0.865) & (0.860) & (0.867) & (0.892) & (0.869) & (0.867) & (0.870) & (0.889) \\
{} & \multirow{2}{*}{105} & 0.818 & 0.813 & 0.821 & 0.871 & 0.871 & 0.869 & 0.874 & 0.902 & 0.886 & 0.885 & 0.889 & 0.908 \\
 &  & (0.829) & (0.823) & (0.833) & (0.873) & (0.878) & (0.874) & (0.880) & (0.905) & (0.890) & (0.889) & (0.892) & (0.910) \\
&&&&&&&&&&&&& \\
\multirow{6}{*}{$\frac{5}{250}$} & \multirow{2}{*}{95} & 1.145 & 1.121 & 1.153 & 1.280 & 1.256 & 1.243 & 1.260 & 1.332 & 1.289 & 1.282 & 1.292 & 1.341 \\
 &  & (1.162) & (1.138) & (1.173) & (1.281) & (1.269) & (1.255) & (1.274) & (1.333) & (1.299) & (1.291) & (1.301) & (1.343) \\
 & \multirow{2}{*}{100} & 1.270 & 1.246 & 1.279 & 1.411 & 1.330 & 1.319 & 1.334 & 1.408 & 1.350 & 1.344 & 1.352 & 1.404 \\
 &  & (1.290) & (1.266) & (1.303) & (1.413) & (1.345) & (1.332) & (1.350) & (1.410) & (1.361) & (1.353) & (1.363) & (1.405) \\
{} & \multirow{2}{*}{105} & 1.243 & 1.227 & 1.249 & 1.379 & 1.350 & 1.341 & 1.354 & 1.429 & 1.383 & 1.378 & 1.385 & 1.438 \\
 &  & (1.267) & (1.250) & (1.277) & (1.380) & (1.366) & (1.356) & (1.371) & (1.430) & (1.395) & (1.388) & (1.397) & (1.439) \\
&&&&&&&&&&&&&\\
\multirow{6}{*}{$\frac{10}{250}$} & \multirow{2}{*}{95} & 1.530 & 1.492 & 1.553 & 1.811 & 1.730 & 1.703 & 1.738 & 1.884 & 1.792 & 1.776 & 1.797 & 1.898 \\
 & & (1.560) & (1.522) & (1.591) & (1.812) & (1.754) & (1.726) & (1.767) & (1.885) & (1.812) & (1.794) & (1.819) & (1.899) \\
 & \multirow{2}{*}{100} & 1.701 & 1.660 & 1.728 & 1.997 & 1.834 & 1.808 & 1.842 & 1.993 & 1.878 & 1.863 & 1.883 & 1.986 \\
 &  & (1.738) & (1.698) & (1.775) & (1.998) & (1.862) & (1.835) & (1.874) & (1.994) & (1.900) & (1.883) & (1.906) & (1.987) \\
{} & \multirow{2}{*}{105} & 1.675 & 1.644 & 1.689 & 1.951 & 1.863 & 1.842 & 1.870 & 2.022 & 1.924 & 1.912 & 1.929 & 2.034 \\
 &  & (1.720) & (1.688) & (1.743) & (1.952) & (1.894) & (1.872) & (1.905) & (2.023) & (1.948) & (1.934) & (1.955) & (2.035)
\end{tabular}
\caption{Exact and approximate square root of mean squared hedging error of the variance-optimal hedge for $\mu=-0.08$, $\sigma = 0.4$, $\mathrm{Skew(X_1)}=\frac{0.1}{\sqrt{250}}$ and varying excess kurtosis $\mathrm{Exkurt(X_1)}$, strike $K$ and maturity $T$; the values in brackets denote the exact and approximate mean squared hedging error of the Black-Scholes hedge with volatility parameter $\sigma = 0.4$}
\label{tab:hedgingError}
}
\end{sidewaystable}

\section{Conclusion} \label{sec:Conclusion}
We provide second-order approximations to the variance-optimal and pure hedge as well to the mean squared hedging errors of these two strategies and the Black-Scholes hedge when the discounted stock price follows a Lévy process and the payoff is smooth. The approximations are obtained by considering the Lévy model of interest as a perturbed Black-Scholes model. More specifically, our approach relies on connecting the Lévy model under consideration with the approximating Black-Scholes model by a curve in the set of stochastic processes. 
The choice of this curve
may be considered as part of the modelling 
as it typically affects the structure 
and possibly even the existence of an approximation. 
We leave a thorough discussion of this aspect and a comparative study to future research.
It any case the curve needs to be chosen such that it leads to
computable expressions that are numerically sufficiently accurate in practically relevant cases. 

Qualitatively, our results show that the deviation of hedges and hedging errors from Black-Scholes is essentially determined by the third and fourth moment of logarithmic returns in the Lévy model and by Black-Scholes sensitivities (cash greeks) of the option. The fine structure of the Lévy process is less relevant. The option contributes to the hedging error primarily through its Black-Scholes gamma.

Quantitatively, for models from the literature and reasonable parameter values, numerical tests indicate that the accuracy of our approximations is excellent for initial capital and hedge ratios and reasonable for their hedging errors. Moreover, our tests suggest that the Black-Scholes strategy is a very good proxy to the variance-optimal one, and its hedging error due to the jumps of the Lévy process is essentially determined by the excess kurtosis of logarithmic stock returns. By comparison with results on discrete-time hedging, one may say that the risk of the Black-Scholes hedge in the presence of jumps is the same as if the Black-Scholes delta is implemented discretely in a Black-Scholes market at time steps
	\begin{equation*} \Delta t = \frac{1}{2} \left(\exkurt{\text{log-returns}}-\left(\skew{\text{log-returns}}\right)^2\right). \end{equation*}

\addcontentsline{toc}{section}{References}

\bibliography{./references}

\end{document}